\documentclass[a4paper,oneside]{article}
\usepackage{comment}
\usepackage{float}

\usepackage[utf8]{inputenc}
\usepackage{alphabeta}

\usepackage{indentfirst}

\usepackage{mathtools}
\usepackage{amsmath}
\usepackage{amssymb}
\usepackage{amsthm}
\usepackage{physics}

\usepackage{setspace}
\RequirePackage{fix-cm}
\usepackage{nicematrix,tikz}

\usepackage{titlesec}
\usepackage{physics}
\usepackage{array}

\usepackage[colorlinks, linkcolor=blue, citecolor=blue]{hyperref}
\usepackage{nameref}
\usepackage{cleveref}

\usepackage{dsfont}
\usepackage{xcolor}
\usepackage{enumitem}
\usepackage{thmtools}

\Crefname{theorem}{Theorem}{Theorems}
\Crefname{lem}{Lemma}{Lemmata}
\Crefname{defn}{Definition}{Definitions}
\Crefname{equation}{Eq.}{Equations}

\newtheorem{theorem}{Theorem}[section]
\newtheorem{prop}[theorem]{Proposition}
\newtheorem{lem}[theorem]{Lemma}
\newtheorem{corollary}[theorem]{Corollary}
\newtheorem{defn}[theorem]{Definition}
\newtheorem{remark}{Remark}

\DeclareMathOperator{\dist}{dist}
\DeclareMathOperator{\supp}{supp}

\DeclarePairedDelimiter\floor{\lfloor}{\rfloor}

\newcommand{\Z}{{\mathbb{Z}}}
\newcommand{\N}{{\mathbb{N}}}
\newcommand{\R}{{\mathbb{R}}}
\newcommand{\C}{{\mathbb{C}}}

\begin{document}

\begin{center}
{\Large {\sc Clustering of higher order connected correlations in C$^*$ dynamical systems}}

\vspace{1cm}

{\em Dedicated to the memory of Petros Meramveliotakis}
\vspace{1cm}

{\large Dimitrios Ampelogiannis and Benjamin Doyon}

\vspace{0.2cm}
Department of Mathematics, King's College London, Strand WC2R 2LS, UK

\end{center}

In the context of $C^*$ dynamical systems, we consider a locally compact group $G$ acting by $^*$-automorphisms on a C$^*$ algebra $\mathfrak{U}$ of observables, and assume  a state of $\mathfrak{U}$ that satisfies the clustering property with respect to a net of group elements of $G$. That is, the two-point connected correlation function vanishes in the limit on the net, when one observable is translated under the group action. Then we show that all higher order connected correlation functions (Ursell functions, or classical cumulants) and all free correlation functions (free cumulants, from free probability) vanish at the same rate in that limit. Additionally, we show that mean clustering, also called ergodicity,  extends to higher order correlations.
We then apply those results to equilibrium states of quantum spin lattice models. Under certain assumptions on the range of the interaction, high temperature Gibbs states are known to be exponentially clustering w.r.t.\ space translations. Combined with the Lieb-Robinson bound, one obtains exponential clustering for space-time translations outside the Lieb-Robinson light-cone. Therefore, by our present results, all the higher order connected and free correlation  functions will vanish exponentially under space-time translations outside the Lieb-Robinson light cone, in high temperature Gibbs states. Another consequence is that their long-time averaging over a space-time ray vanishes for almost every ray velocity.

\tableofcontents

\section{Introduction}
In this paper we are concerned with the clustering properties of multi-point connected correlations, or joint cumulants, of states of $C^*$ algebras, with respect to group actions. We show that if the 2-point connected correlation (covariance) between two observables vanishes when we move one ``infinitely far'', under the group action, then the $n$-th cumulant between $n$ observables also vanishes at the same rate. Additionally, if we average over the group action and the 2-point connected correlation vanishes in the mean, then the $n$-point connected correlation also vanishes in the mean. We show these both for classical cumulants, or Ursell functions, and free cumulants from the theory of free probability \cite{speicher_2019_notes}.

These results are particularly interesting in quantum many-body physics, for the group being that of translations along space and / or time. For instance, this includes quantum spin lattice (QSL) models with translation symmetry (such as the $\mathbb Z^d$ lattice). Indeed, in this context higher-order correlations are relevant as they characterise higher order hydrodynamics \cite{myers_2020_fluctuations,doyon_2020_fluctuations} and in particular provide bounds on diffusion coefficients \cite{doyon_2019_diffusion}.
Higher-order correlations also serve as a measure of chaos in quantum systems via out-of-time-order correlators \cite{garcia_2022_OTOC_chaos_review,bhattacharyya_2022_quantum_chaos}, and as a measure of multi-partite entanglement \cite{zhou_2006_multiparty,pappalardi_2017_multi_entanglement}. The accuracy of the mean field approximation depends on the assumption that correlations between observables are negligible, and to improve it higher order cumulants can be used \cite{Fowler_2023_cumulant_expansion, Kramer_2015_generalisedMFT, Robicheaux_2021_beyond}, generally by a cumulant expansion method \cite{kubo_1962_cumulant, fricke_transport_1996, Kira_2008_clusterexp, Sanchez_2020_cumulant}. Our result on vanishing in the mean is especially relevant to averages along rays in space-time, as there we have recently proven the vanishing in the mean of the covariance within QSL with a large amount of generality \cite{ampelogiannis_2023_almost}. The vanishing of covariance was a crucial ingredient in the proof of the hydrodynamic projection principle for two-point functions \cite{doyon_hydrodynamic_2022,ampelogiannis_long-time_2023}, and we expect the projection principle for higher-point functions \cite{doyon_2023_mft} can be established from our present results.

The connected correlations generally considered in the above applications are the classical cumulants, which measure classical independence (as random variables) between observables. However, one can also consider different types of independence. In non-commutative probability one generally considers the notion of ``freeness'', introduced by Voiculescu who developed the theory \cite{voiculescu_1985_Symmetries,voiculescu_1986_addition,voiculescu_1987_multiplication}, and the free cumulants measure free independence, introduced by Speicher \cite{speicher_1994_multiplicative}. Recently a connection between free probability theory and the  Eigenstate Thermalization Hypothesis (ETH) was made in \cite{Pappalardi_2022_ETH_free}, where higher order correlation functions are determined by the free cumulants. Additionaly, in \cite{jindal_2024_free} quantum chaos and the decay of the out-of-time-order correlators are studied using free cumulants.  In light of this recent activity using tools from free probability in quantum statistical mechanics, we consider here both classical and free cumulants. 

The vanishing of higher-order correlation functions has been considered using cumulant expansion methods for classical and quantum  systems \cite{Ueltschi2003Cluster}. A similar result in product states of finite-volume systems extending the Lieb-Robinson bound on $n$-partite correlations was shown in \cite{Tran_2017_Lieb_Robinson_npartite}. However, we are not aware of results as general as those established here, in particular for free cumulants.

This paper is organised as follows. In \Cref{section:set-up} we discuss the precise set-up of C$^*$ dynamical systems in which we work and then in \Cref{section:cumulants} we precisely define both the classical (connected correlations) and the free  cumulants  for a state of a C$^*$algebra. In \Cref{section:results} we provide our main results, with rigorous statements. In \Cref{section:qsl}
we apply our results to quantum lattice models with either short-range or long-range interactions. Finally, the proofs of the theorems discussed in \Cref{section:results} are done in \Cref{section:proofs}.

\section{Set-up} \label{section:set-up}
We consider a unital $C^*$-algebra $\mathfrak{U}$, as the set of observables of our physical system, and a group $G$ that acts on $\mathfrak{U}$ by $*$-automorphisms $α_g$, $g \in G$. The physical system in mind can be classical ($\mathfrak{U}$ abelian) or quantum  ($\mathfrak{U}$ non-commutative). We only make the assumption that the group $G$ is locally compact; it can for example describe time evolution, $G=\R$, or space translations, $G=\Z^D$ for a spin lattice and $G=\R^D$ for continuous systems in $D$ dimensions.

\begin{defn}[$C^*$  Dynamical System] \label{defn:dynamicalsystem}
A  $C^*$ dynamical system is a triple $(\mathfrak{U},G,α)$ where $\mathfrak{U}$ is a unital $C^*$-algebra, $G$ a locally compact group, and $α$ is a representation of the group $G$ by $^*$-automorphisms of $\mathfrak U$.
\end{defn} 

We examine states $ω$ of $\mathfrak{U}$ that are clustering with respect to $G$, in the sense that the connected correlation between two observables vanishes under some limit action, e.g.\ space-translating one observable infinitely far.

\begin{defn}[Clustering] \label{defn:clustering}
A state $ω$ of a $C^*$ dynamical system  $(\mathfrak{U},G,α)$ is called $G$-clustering if there exists a net $ g_i \in G$ such that 
\begin{equation}\label{clustering}
    \lim_i \big(ω( α_{g^i}(A) B) - ω(α_{g^i}A) ω(B) \big)= 0 , \ \forall A,B \in \mathfrak{U}.
\end{equation} 

\end{defn}

The decay of 2-point connected correlations, for space-translations, has been established rigorously in many physical systems. In classical systems see \cite{ruelle_statistical_1974}; in particular a one-dimensional classical lattice gas satisfies the clustering property, as shown by Ruelle  \cite{Ruelle_1968_classical_gas}. For quantum spin chains (QSC), Araki \cite{Araki:1969bj} showed exponential clustering in Gibbs states, for finite-range interactions. In QSC with short-range interactions the correlations also decay exponentially in Gibbs states \cite{perez_2023_locality}. Clustering for quantum gases was established by Ginibre \cite{Gibibre_1965_quant_gases}. Cluster expansion methods can be used quite generally, to show decay of correlations for both classical and quantum systems \cite{park_cluster_1982,Ueltschi2003Cluster}.

In the non-commutative setup, it can be the case that the algebra is asymptotically abelian under the action of $G$, i.e.\ there exists a net $g_i \in G$:
\begin{equation}
    \lim_i \norm{ [ α_{g^i} A, B] }=0 , \ \forall A,B \in \mathfrak{U}.
\end{equation}
It follows from asymptotic abelinanness that every factor state will satisfy the cluster property \Cref{clustering} \cite[Example 4.3.24]{bratteli_operator_1987}.  A physically relevant example of factors is that of thermal equilibrium states, described by the Kubo-Martin-Schwinger (KMS) condition \cite[Section 5.3]{bratteli_operator_1997}. An extremal KMS state is factor \cite[Theorem 5.3.30]{bratteli_operator_1997}, hence it has clustering properties. 

\begin{prop}
    Consider a $C^*$ dynamical system  $(\mathfrak{U},G,α)$ that is asymptotically abelian. If $ω$ is a factor state, then ω is $G$-clustering. 
\end{prop}

\begin{corollary}
Consider a $C^*$ dynamical system  $(\mathfrak{U},G,α)$ that is asymptotically abelian and $τ_t$ is the strongly continuous group of time evolution. If $ω_β$ is an extremal $(τ,β)$-KMS state, then $ω$ is factor and hence $G$-clustering.
\end{corollary}

A concrete example is a quantum spin chain, such as the XXZ-chain: the spin chain is asymptotically abelian under space translations $ \lim_{x\to \infty} \norm{ [ α_x A, B] }=0$ \cite[Example 4.3.26]{bratteli_operator_1987} and at non-zero temperature there is a unique KMS state \cite{Araki1975uniqueness} which is subsequently factor and hence clustering with respect to space translations. A continuous example is the ideal Bose gas, described by the CCR (canonical commutation relations) algebra, which is also asymptotically abelian for space translations \cite[Example 5.2.19]{bratteli_operator_1997}. The equilibrium state of the ideal Bose gas is clustering for both space and time translations, in the single phase region \cite[Section 5.2.5]{bratteli_operator_1997}. Of course, asymptotic abelianness is not a neccessary condition, the  ideal Fermi gas is described by the CAR (canonical anticommutation relations) algebra which is not asymptotically abelian \cite[Section 5.2.21]{bratteli_operator_1997}, but its equilibrium states are clustering both in space and in time \cite[Section 5.2.4]{bratteli_operator_1997}.

\section{Higher order cumulants and free cumulants} \label{section:cumulants}

\subsection{Connected correlations -- classical cumulants} \label{section:classical_cumulants}
Consider a $C^*$ algebra $\mathfrak{U}$ and a state $ω$. We start by defining the  connected correlations between $n$-observables, or joint classical cumulants, for the state $ω$. The first three cumulants are 
\begin{equation} \label{eq:example_cumulants}
 \begin{array}{*3{>{\displaystyle}lc}p{5cm}}
  c_1 ( A_1 )  &=& ω(A_1) \\
  c_2( A_1, A_2) &=& ω(A_1 A_2)-ω( A_1) ω(A_2)\\
   c_3(A_1, A_2, A_3) &=& ω(A_1 A_2 A_3) - ω(A_1)ω(A_2 A_3) - ω(A_2) ω(A_1 A_3)\\&&
   - ω(A_3) ω(A_1A_2) +2 ω(A_1)ω(A_2)ω(A_3) 
        \end{array}    
\end{equation}
for $A_1,A_2,A_3 \in \mathfrak{U}$. To give a general definition, we have to keep in mind that the observables might not commute. We define for every $n\in \N$ the functional
\begin{equation}
    ω_n( A_1, A_2, \cdots , A_n) \coloneqq ω(A_1 A_2 \cdots A_n)  \ , \ \ A_1,\cdots , A_n \in \mathfrak{U}.
\end{equation}
Then, we extend it to the multiplicative family of moments $(ω_π)_{π \in P}$, where $P = \cup_{n \in \N} P(n)$ and $P(n)$ the set of all partitions of $\{1,2,\cdots,n\}$, as follows:
\begin{equation}
    ω_π (A_1, A_2, \cdots, A_n)  \coloneqq \prod_{V \in π} ω_{|V|} (A_1, A_2, \cdots , A_n | V)
\end{equation}
where $ ω_s(A_1, A_2, \cdots , A_n | V)$ keeps the indices $i\in V$ in the correct order; for any $s\in \N$, $i_1,\cdots , i_s\in \{1,\cdots,n\}$ with $i_1 < \cdots <i_s$ and $V=\{i_1,\cdots ,i_s\}$:
\begin{equation} \label{eq:notation_ordering}
    ω_s(A_1, A_2, \cdots , A_n | V)  \coloneqq ω_s ( A_{i_1}, \cdots , A_{i_s} ).
\end{equation}
The most direct way of defining the classical cumulants of the state $ω$ is by a sum, over all partitions, of products of joint moments \cite{Speed_1983_Cumulants1}:
\begin{defn} \label{defn:classical_cumulants}
  Consider  a C$^*$ algebra  $\mathfrak{U}$ and a state $ω$. The classical cumulants of $ω$  are defined by 
  \begin{eqnarray}
    c_n (A_1, A_2,\ldots,A_n) &=& \sum_{π \in P(n)} (-1)^{|π|-1}( |π| - 1)! \prod_{V \in π} ω_{|V|} (A_1, A_2, \cdots , A_n | V)\nonumber \\
    &=&\sum_{π \in P(n)} (-1)^{|π|-1}( |π| - 1)! \, ω_π( A_1 , A_2, \ldots , A_n)
\end{eqnarray}
for $n \in \N$ and $A_1, A_2, \ldots, A_n\in\mathfrak U$.
\end{defn}

A more useful approach, is an indirect recursive definition as multilinear functionals $c_n: \mathfrak{U}^n \to \C$ that we extend to a multiplicative family $\{c_π\}_{π \in P}$:
\begin{equation}
    c_π(A_1,A_2,\cdots, A_n) = \prod_{V\in π} c_{|V|} (A_1,A_2,\cdots,A_n | V)
\end{equation}
and require that they satisfy the moments-to-cumulants formula
\begin{equation} \label{eq:moments_to_cumulants}
    ω_n(A_1,A_2, \ldots, A_n) = \sum_{π \in P(n)} c_π (A_1,\cdots,A_n).
\end{equation}

For our proofs, it will be important to use the following recursive definition of the set $P(n)$, where the recursion involves the possible sets in which the particular element $1$ lies (similarly, any other element could have been chosen):
\begin{equation}\label{eq:recursionP}
    P(n) = \{ \{ V\} \cup \pi: V\ni 1,\pi \in P(\{1,\ldots,n\}\setminus V)\}
\end{equation}
where we use the notation $P(V)$ to denote the set of all partitions of the elements of the set $V$ (in particular $P(n) = P(\{1,\ldots,n\})$).

\subsection{Free cumulants} \label{section:free_cumulants}
A double $(\mathfrak{U},ω)$ consisting of a non-commutative C$^*$ algebra $\mathfrak{U}$ and a state $ω$ forms a non-commutative probability space  \cite{speicher_2019_notes}. The free cumulants where introduced by Speicher \cite{speicher_1994_multiplicative} in the context of free probability theory. The definition is based on the non-crossing partitions, which are partitions with blocks that do not cross, in a natural diagrammatic representation where elements are ordered (say along a circle). We denote by $NC(n)$ the set of all non-crossing partitions of $\{1,2,\ldots, n\}$ and $NC= \cup_{n\in\mathbb N} NC(n)$. We give a recursive definition via the moments-to-cumulants formula, based on the discussion in \cite{speicher_2019_notes}:

\begin{defn}[Free cumulants]\label{defn:free_cumulant}
    Consider a $C^*$ algebra $\mathfrak{U}$ and a state $ω$. The free cumulants are a multiplicative family $κ_π$ over $π\in NC$, in the sense that $κ_n:\mathfrak{U}^n \to \C$ are multilinear for all $n \in \N$ and (using a notation as in Eq.~\eqref{eq:notation_ordering})
    \begin{equation} \label{eq:multiplicative_free}
    κ_π(A_1,A_2,\ldots, A_n) = \prod_{V\in π} κ_{|V|} (A_1,A_2,\ldots,A_n | V)
\end{equation}
for $n\in \N$, $π \in NC(n)$ and $A_1,A_2,\ldots,A_n \in \mathfrak{U}$, such that they satisfy
\begin{equation} \label{eq:moments_to_free_cumulants}
    ω_n(A_1,A_2, \ldots, A_n) = \sum_{π \in NC(n)} κ_π (A_1,\ldots,A_n).
\end{equation}
The $n$-th free cumulant corresponds to the maximal partition $κ_n \coloneqq κ_{\mathds{1}_n}$.
\end{defn}

See also \Cref{appendix:mobius} for a more direct (equivalent) definition. Note that the first three free cumulants are the same as the classical ones \Cref{eq:example_cumulants}, as all the partitions of $\{1,2\}$ and $\{1,2,3\}$ are non-crossing. For $n=4$ there is one crossing partition, $\{ \{1,3\}, \{2,4\} \}$, and hence for $n\geq 4$ the free cumulants are different from the classical ones.

Again, for our proofs, it will be important to use a recursive definition of the set of non-crossing partitions. Let us denote $NC(V)$ the non-crossing partitions of the set $V$, and in particular $NC(n) = NC(\{1,\ldots,n\})$. For a subset $V$ of $\{1,\ldots,n\}$, let us denote $\pi^V$ the unique partition of $V$ that is the minimal one (with the smallest number of parts) such that every part is composed of consecutive elements: every $W\in \pi^V$ is of the form $\{w_1,w_2,\ldots\}$ with $w_j = a+j\,{\rm mod}\,n$ for some $a$. Then we have
\begin{equation}\label{eq:recursionNC}
    NC(n) = \{ \{ V\} \cup \cup_j\pi_j: V\ni 1, \pi_j \in NC(W_j) \mbox{ for } \pi^{\{1,\ldots,n\}\setminus V}=\{W_1,W_2,\ldots\}\}
\end{equation}
and similarly, any other element instead of 1 could have been chosen to write the recursion. Note the difference with \eqref{eq:recursionP}: the non-crossing condition is implemented in the smaller set of partitions that combine with $V$, i.e.~those of the form $\cup_j \pi_j$: restricting to sets formed of consecutive elements of $\{1,\ldots,n\}\setminus V$ is what imposes the non-crossing condition.

\section{Main results: Clustering for $n$-th order cumulants}  \label{section:results}

Our main result is clustering for the $n$-th classical, and free, cumulants in any state that is clustering for the covariance (second cumulant), as per \Cref{defn:clustering}. In fact, all higher cumulants inherit the same rate of decay as that of the covariance.  A weaker form of clustering, ergodicity, can also be extended to higher cumulants; averaging the cumulants over the group action (e.g.\ time average) gives $0$. 

\subsection{Clustering for $n$-th order (free) cumulants} \label{section:results1}

Taking advantage of the moment-to-cumulants formula \Cref{eq:moments_to_cumulants} we can inductively show that whenever the two-point connected correlation vanishes under some limit of the group action, $ \lim_{i}  \big(ω( \alpha_{g^i} (A) B) -  ω(\alpha_{g^i} A) ω(B)\big) = 0$, then all higher cumulants will also vanish at the same limit. We show this in detail in \Cref{section:proof_1} for the free cumulants, but the proof for the classical ones is the same.
\begin{theorem} \label{th:cumulants_general}
    Let $(\mathfrak{U},G,α)$ be a C$^*$  dynamical system and $ω$ a $G$-clustering state, i.e.\  there exists a net $g^i \in G$ such that 
    \begin{equation} \label{eq:clustering_th1}
       \lim_{i}  \big(ω( α_{g^i} (A) B) -  ω(α_{g^i} A) ω(B) \big) = 0,   \ \forall A,B \in \mathfrak{U}
    \end{equation}
then for any $n\in \N$ and $A_1,A_2,\ldots , A_n \in \mathfrak{U}$ the free cumulant  vanishes 
\begin{equation} \label{eq:main}
    \lim_{i} κ_n( \alpha_{g^i}A_1 , A_2, \ldots, A_n)
    =
    \lim_{i} κ_n( A_1 , A_2, \ldots, \alpha_{g^i} A_n)= 0 .
\end{equation}
The same holds for the classical cumulants $c_n$:
\begin{equation} \label{eq:maincumu}
    \lim_{i} c_n( \alpha_{g^i}A_1 , A_2, \ldots, A_n) =
    \lim_{i} c_n( A_1 , A_2, \ldots, \alpha_{g^i}A_n)= 0 .
\end{equation}
\end{theorem}

The theorem also holds for states that are clustering only for a $^*$-subalgebra of observables. Suppose   $\mathfrak{V} \subset \mathfrak{U}$ is a  $^*$-subalgebra and that 2-point clustering, \Cref{eq:clustering_th1}, holds for all $A,B \in \mathfrak{V}$. Then the $n$-th cumulants between any $A_1,A_2,\ldots,A_n \in \mathfrak{V}$ satisfy the clustering property.

Note that under the assumptions of \Cref{th:cumulants_general} it is not necessarily true that 
\begin{equation}\label{eq:cumulant_i}
 \lim_{i} κ_n( A_1 ,  \ldots, \alpha_{g^i} (A_m), \ldots, A_n) = 0 
\end{equation}
for $1<m<n$ (similarly for the classical cumulants). This is because a clustering state, as per \Cref{defn:clustering}, does not necessarily satisfy the three-element clustering property $ \lim_{i}  \big(ω(  A \alpha_{g^i}(B) C) -  ω(\alpha_{g^i} B) ω(AC)\big) = 0$. However, if we add this property as an assumption, then \Cref{th:cumulants_general} can be extended so that \Cref{eq:cumulant_i} holds for any $m=1,2,\ldots,n$. This is shown in \Cref{section:proof_1}, at the end of the proof. Note that in asymptotically abelian algebras the two element clustering property is equivalent to the three-element property \cite[Theorems 4.3.22 \& 4.3.23]{bratteli_operator_1987}, hence all factor states satisfy the more general case.

We can also have a set of observables $\{ α_{g^i}  A_1,  α_{g^i} A_2, \ldots ,  α_{g^i} A_m\}$ translated away from another set $\{ A_{m+1}, \ldots , A_n\}$, this is shown in \Cref{section:proof_3}.

\begin{theorem} \label{th:clustering_for_groups}
Consider the assumptions of \Cref{th:cumulants_general}. For any $m < n \in N$ and $A_1, A_2 , \ldots, A_n \in \mathfrak{U}$ it follows that
\begin{equation}
  \lim_i  κ_n( α_{g^i}  A_1,  α_{g^i} A_2, \ldots ,  α_{g^i} A_m, A_{m+1}, \ldots , A_n ) =0.
\end{equation}
The same true is for classical cumulants $c_n$. 
\end{theorem}
In the results that follow, we only consider the simplest case of translating one element under the group action and examining the clustering properties of joint cumulants. However, using the same techniques, these results can be generalised in the same manner as \Cref{th:clustering_for_groups}.

\subsection{Rate of decay}
Consider a clustering state $ω$ of a $C^*$ dynamical system $(\mathfrak{U},G,α)$ and a function $f:G \to \R^+$. We call $ω$ $f(g)$-clustering if the rate of decay of the covariance is $1/f(g)$ in some $^*$-subalgebra $\mathfrak{V} \subset \mathfrak{U}$. For example, KMS states in quantum lattice models are exponentially clustering in space, that is $e^{\lambda |x|}$-clustering for some $\lambda>0$ \cite{Araki:1969bj}. The proof of \Cref{th:cumulants_general} can easily be modified to show that for a $f(g)$-clustering state the $n$-th classical, and free, cumulant also decays with rate $1/f(g)$, for all $n \geq 2$. Both proofs are done in \Cref{section:proof_1}.

\begin{theorem} \label{th:decay}
    Let $(\mathfrak{U},G,α)$ be a $C^*$  dynamical system and $ω$ a f(g)-clustering state for a $*$-subalgebra $\mathfrak{V} \subset \mathfrak{U}$ , i.e.\  there exists a net $g^i \in G$ such that
    \begin{equation}
       \lim_{i}  f(g^i)\big(ω( α_{g^i}(A) B) - ω( α_{g^i}A) ω(B) \big)= 0 , \ \forall A,B \in \mathfrak{B} 
    \end{equation}
for some non-zero $f: G \to \R^+$. Then for any $n\in \N$ the free cumulants vanish at the same rate:
\begin{equation} \label{eq:maindecay}
    \lim_{i} f(g^i)κ_n( α_{g^i}A_1 , A_2, \cdots, A_n) = 0 , \ \forall A_1, A_2, \ldots, A_n \in \mathfrak{B}. 
\end{equation}
The same holds for the classical cumulants $c_n$.
\end{theorem}

Note that the theorem  includes the case $\mathfrak{V} = \mathfrak{U}$, but does not require that $\mathfrak{V} $ is norm closed. 

Additionally, bounds on the second cumulant also apply to higher order cumulants. In a similar manner one can show that $|ω( α_{g}(A) B) - ω( α_{g}A) ω(B)| \leq C_2\norm{A} \norm{B}f(g)$ implies that $|κ_n( α_{g}A_1 , A_2, \cdots, A_n)| \leq  C_n \prod_j \norm{A_j} f(g)$ where $C_n>0$. We obtain such a result for quantum lattice models in Section \cref{subsection:52}.

\subsection{Mean Clustering}
We can go further and assume a weaker clustering property called mean clustering, a form of ergodicity with respect to group actions on $C^*$-algebras \cite[Section 4.3]{bratteli_operator_1997}. That is, instead of clustering states, with covariance that vanishes, we consider states such that the mean of the covariance over the group action goes to $0$. This is relevant with our recent results of almost everywhere ergodicity \cite{ampelogiannis_2023_almost}, that we discuss in \Cref{section:qsl}.

In general, it is not always possible to define an invariant mean over the group action.
This is possible in groups called amenable \cite{greenleaf_1969_invariant}. If $μ$ is the Haar measure of the locally compact group $G$, then one of the equivalent definitions of an amenable group is that for every compact $K\subset G$ there exists a net $U_i \subset G$, with $μ(U_i) \leq \infty$, such that
\begin{equation} \label{eq:amenable}
    μ( U_i Δ gU_i)/ μ(U_i) \to 0 , \ \forall g \in K
\end{equation}
where $gU_i = \{ h \in G : h= g u, u\in U_i \}$ is $U_i$ translated by $g$, and $A Δ B = (A \cup B) \setminus(A \cap B)$. Then, one can define an invariant mean for functions over $G$  as
\begin{equation}
    M(f) = \lim_i \frac{1}{μ(U_i)} \int_{U_i} f(g) \,dμ(g).
\end{equation}
Every locally compact abelian group is amenable and we restrict the next theorem to such groups, but locally compact amenable would still be sufficient. Time and space translations are of course described by amenable groups, such as  $G=\R^D$ which clearly satisfies \Cref{eq:amenable} for a net of balls of increasing radius.
 
\begin{theorem} \label{th:ergodicity}
    Let $(\mathfrak{U},G,α)$ be a $C^*$ dynamical system and $G$ a locally compact abelian group with Haar measure $μ$. Let $ω$ be a state such that there exists a net $U_i \subset G$ with
    \begin{equation} \label{eq:ergodicity_general}
        \lim_i \frac{1}{μ(U_i)} \int_{U_i} \big( ω( α_{g}(A) B) - ω( α_{g}A) ω(B)  \big) \,dμ(g) =0, \forall A,B \in \mathfrak{U}.
    \end{equation}
It follows that the free cumulants also have vanishing mean:
\begin{equation}
     \lim_i \frac{1}{μ(U_i)} \int_{U_i} κ_n( α_{g}A_1 , A_2, \cdots, A_n) \,dμ(g)= 0 
\end{equation}
for all $A_1,A_2,\ldots,A_n\in\mathfrak{U}$. The same holds for the classical cumulants $c_n$.
\end{theorem}

In the context of ergodicity the state $ω$ is generally considered to be $G$-invariant, that is $ω(α_gA)=ω(A)$. However, it is not necessary to assume invariance in order to prove \Cref{th:ergodicity}, so we have kept the expression as general as possible.  The proof of \Cref{th:ergodicity} is done in Section \ref{section:proof_2}.

\subsection{Banach Limit clustering}

An interesting observation is that the proofs only rely on the linearity of the limit. This allows us to consider generalised limits, such as a Banach limit, in order to extend our theorems so that whenever a Banach limit of the second cumulant vanishes, then the $n$-th cumulants also vanish for the same limit.

\begin{theorem}
    Let $(\mathfrak{U},G,α)$ be a dynamical system and $G$ a locally compactgroup . Let $ω$ be a state such that there exists a net $g_i \in G$ and a Banach Limit $\widetilde \lim$ with
    \begin{equation} \label{eq:7 }
        \widetilde{\lim_i}\,  \big( ω( α_{g^i}(A) B) - ω( α_{g}A) ω(B)  \big)  =0, \ \forall A,B \in \mathfrak{U}.
    \end{equation}
It follows that the free cumulants also vanish at the same limit
\begin{equation}
    \widetilde {\lim_i}\, κ_n( α_{g^i}A_1 , A_2, \cdots, A_n)  = 0, \ \forall A_1,A_2,\ldots,A_n \in \mathfrak{U}.
\end{equation}
The same holds for the classical cumulants.
\end{theorem}

\section{Application to QSL: space-like clustering from Lieb-Robinson bound} \label{section:qsl}
We consider a quantum spin lattice (QSL)  with either short-range interactions (exponentially decaying or finite range) or long-range (power-law decaying). The $C^*$-algebra of observables is a quasi-local algebra $\mathfrak{U}= \overline{\cup_{Λ\subset \Z^D} \mathfrak{U}_{Λ}}$ over $\Z^D$ and we have the groups of space translations $ι_x$ and time evolution $τ_t$, acting as $^*$-automorphisms on $\mathfrak{U}$. We call the triple $(\mathfrak{U},ι,τ)$ a quantum lattice $C^*$ dynamical system. See \cite[Section 6.2]{bratteli_operator_1997} for the detailed construction. Note that existence of the infinite volume dynamics for long-range interactions has been established, see \cite{nachtergaele_quasi-locality_2019}.

A central result in the context of QSL is the Lieb-Robinson bound, which  exponentially bounds the effect of any time-evolved observable outside a light-cone. The Lieb-Robinson bound was initially shown for finite range interactions \cite{Lieb:1972wy} and later extended to short range  interactions, see for example \cite{nachtergaele_2006_LR}. In the case of long-range interactions, the bound was extended by Hastings and Koma \cite{Hastings_2006_spectralgap}, but with a light-cone velocity that diverges with distance. However, in \cite{Chen_2019_finite_scrambling,Kuwahara_2020_linear_LR}, under the assumption of a power-law decaying interaction $\sim 1/r^a$ with $a>2D+1$, where $D$ is the lattice dimension, it is shown that one obtains a linear light-cone. Here, we are not interested in the specifics on the Lieb-Robinson bounds, but on the existance of the linear light cone and the resulting asymptotic abelianness for space-time translations outside of it. To summarise, for short-range interactions or power-law decaying with exponent  $a>2D+1$, there exists a $υ_{LR}>0$, called the Lieb-Robinson velocity, and a $λ>0$ such that for any local observables $A,B$ 
\begin{equation}  
      \norm{[ τ_t (A) , B]} \leq \begin{cases}
 L_{A,B} \exp\{-λ( \dist ( A,B) - υ_{LR}|t|)\},  &\text{for short-range}\\
 \\[1pt]
\displaystyle L^{\prime}_{A,B} \frac{ |t|^{D+1} }{(\dist (A,B) -υ_{LR}|t|)^{a-D} }, &\text{for long-range}
\end{cases}\label{eq:liebrobinsonbound}
\end{equation}
where $\dist(A,B)$ is the distance between their supports and $L_{A,B}$, $L^{\prime}_{A,B}$ depend on the norms and support sizes of $A,B$. 

This section is divided into three subsections. In \ref{susbsection:51} we apply results from the previous section to obtain clustering and exponential clustering, for higher order cumulants, for space-time translations outside the Lieb-Robinson light cone, in physically relevant states. We also discuss ergodicity results for higher order connected correlations, applying results from \cite{ampelogiannis_2023_almost}.  In \ref{subsection:52} we consider an inverse power-law bound on second order connected correlations and obtain a bound for $n$-th order connected correlations $c_n(A_1(x_1,t_1),\ldots, A_n(x_n,t_n))$ at different (not necessarily ordered) times. This bound requires a three element clustering property, for $ω \big( A B(x,t) C) - ω(AC) ω(B(x,t))$, which comes as a consequence of the Lieb-Robinson bound in states that satisfy the usual two element clustering for $ω(A(x,t) B) -ω(A(x,t))ω(B)$. This is discussed in \ref{section:three-element-clustering}. These results are complemented by Appendices \ref{appendix:lemma} and \ref{appendix:space-like}.

\subsection{Clustering for space-like translations and almost everywhere ergodicity} \label{susbsection:51}

The Lieb-Robinson bound  implies that $\mathfrak{U}$ is asymptotically abelian for space-like translations \cite[Theorem 4.2]{ampelogiannis_2023_almost}:
\begin{equation}
    \lim_{x \to \infty} \norm{ [ι_x τ_{xυ^{-1}}A,B]} \to 0     , \ υ> υ_{LR}, \ A,B \in \mathfrak{U}.
\end{equation}
Therefore one concludes that factor states are clustering with respect to space-time translations outside the Lieb-Robinson light-cone:
\begin{equation}
    \lim_{x \to \infty} \big( ω( ι_x τ_{xυ^{-1}}A B) - ω(A) ω(B) \big) =0 , \ υ> υ_{LR}, \ A,B \in \mathfrak{U}.
\end{equation} 
This is shown in detail in \cite{ampelogiannis_2023_almost} for factor invariant states and short-range interactions, but generalising the proof to factor states and long-range is immediate. Therefore, by \Cref{th:cumulants_general} the higher order classical and free cumulants also satisfy the clustering property in factor states:
\begin{corollary} \label{eq:qsl_clustering}
Consider a quantum lattice $C^*$ dynamical system $(\mathfrak{U},ι,τ)$ with either short-range (finite range or exponentially decaying) or power-law decaying ($a>2D+1$)  interaction and a factor state $ω$ of $\mathfrak{U}$. It follows,  by  \cite[Theorem 4.2]{ampelogiannis_2023_almost}, that $ω$ is space-like clustering 
and consequently the $n$-th free and classical cumulants are also space-like clustering:  
    \begin{equation}
    \lim_{t \to \infty} c_n ( ι_{\floor{ \boldsymbol{v}t}}τ_t A_1, A_2, \ldots, A_n)  =  0 , \ \forall |\boldsymbol{v}| > υ_{LR}  , \forall A_1,A_2,\ldots,A_n \in \mathfrak{U}
\end{equation}
\end{corollary}

Additionally, in the case of short-range interactions, if we consider exponentially space-clustering states, such as high temperature KMS states, we can combine the exponential bound on the commutator with  space-clustering to obtain clustering for space-like translations, as shown in \cite[Theorem C.1]{ampelogiannis_long-time_2023}. Thus, by \Cref{th:decay}, we ultimately get exponential space-like clustering for all cumulants in the short-range case.

\begin{corollary} \label{th:space-like-clustering}
Consider a quantum lattice $C^*$ dynamical system $(\mathfrak{U},ι,τ)$ with short-range (finite range or exponentially decaying) interaction and an exponentially space clustering  state $ω$ of $\mathfrak{U}$. It follows, by \cite[Theorem C.1]{ampelogiannis_long-time_2023}, that $ω$ is exponentially space-like clustering for some $λ>0$:
\begin{equation}
    \lim_{t \to \infty} e^{λt} \big( ω\big( ι_{\floor{ \boldsymbol{v}t}}τ_t(A) B \big) - ω(ι_{\floor{ \boldsymbol{v}t}}τ_tA) ω(B) \big) = 0 , \forall|\boldsymbol{v}|> υ_{LR} ,\ A,B \in \mathfrak{U}_{\rm loc}
\end{equation}
Consequently, all $n$-th cumulants are clustering w.r.t space-like translations:
\begin{equation}
    \lim_{t \to \infty}e^{λt} c_n (  ι_{\floor{ \boldsymbol{v}t}}τ_tA_1, A_2, \ldots, A_n)  = 0  , \forall|\boldsymbol{v}|> υ_{LR} 
\end{equation}
for all $A_1, A_2,\ldots , A_n \in \mathfrak{U}_{\rm loc}$, and the same is true for the free cumulants.
\end{corollary}

A weaker form of clustering, ergodicity, can also be useful in the study of quantum lattice models. A recent result in (short-range) quantum lattice models is almost-everywhere ergodicity \cite{ampelogiannis_2023_almost}, which shows that for almost every $υ \in \R$ the long-time averaging of connected correlations over a space-time ray vanishes,
\begin{equation}
    \lim_{T \to \infty} \frac{1}{T} \int_0^T  ω \bigg( ι_{\floor{ \boldsymbol{v}t}}τ_t (A) B \bigg)  \,dt = ω(A) ω(B). \label{eq:maintheorem}
\end{equation}
Combining this result with \Cref{th:ergodicity} we show:

\begin{corollary} \label{th:almost-everywhere-clustering}
Consider a quantum lattice $C^*$ dynamical system $(\mathfrak{U},ι,τ)$ with short-range (finite range or exponentially decaying) translation invariant interactions  and a factor $ι,τ$-invariant state $ω$ of $\mathfrak{U}$. For almost every $υ\in \R$, and every lattice direction $\boldsymbol{q}= \frac{\boldsymbol{x}}{|\boldsymbol{x}|}$, $\boldsymbol{x}\in \Z^D$, the long-time averaging of any $n$-th cumulant (or any $n$-th free cumulant), over a space-time ray with velocity $\boldsymbol{v}=υ \boldsymbol{q}$, will vanish:
\begin{equation}
    \lim_{T \to \infty} \frac{1}{T} \int_0^T  c_n \bigg( ι_{\floor{ \boldsymbol{v}t}}τ_t A_1, A_2,\ldots,  A_n \bigg)  \,dt = 0.
\end{equation}
for all $A_1, A_2, \ldots, A_n \in \mathfrak{U}$. 
\end{corollary} 

A few remarks are in order as to which physically relevant states satisfy the above  corollaries \ref{eq:qsl_clustering}, \ref{th:space-like-clustering}. In short-range interaction quantum spin chains, $D=1$, we have two cases. For finite range interactions any KMS state at non-zero temperature satisfies exponential clustering, by the result of Araki \cite{Araki:1969bj}. For exponentially decaying interactions Araki's result has been extended, but only for high enough temperature \cite{perez_2023_locality}. At higher dimensions, it is shown that high temperature KMS states  are exponentially clustering \cite{frohlich_properties_2015}. These cases fall into the single phase regimes of QSL and will satisfy Corollary \ref{th:space-like-clustering}. For lower temperatures, where multiple phases exist\footnote{The KMS state is not unique, but the KMS states of fixed temperature form a convex set.}, clustering, not necessarily exponential, still holds in space-invariant extremal KMS states, as these are factor \cite[Theorem 5.3.30]{bratteli_operator_1997}, hence these will satisfy \Cref{eq:qsl_clustering}. Note that Corollaries \ref{eq:qsl_clustering}, \ref{th:space-like-clustering} hold for interactions that are not necessarily translation invariant, while  Corollary \ref{th:almost-everywhere-clustering} requires translation invariance and that the state is space and time invariant, which is automatically true for KMS states  in the single phase regime and otherwise assumed at lower temperatures.

\subsection{Bound on $n-$th order connected correlations} \label{subsection:52}

In applications, it might be useful to have a more explicit expression of clustering, in terms of a bound. For example, see \cite{doyon_2019_diffusion}, where a Lieb-Robinson type bound on the $n$-th order cumulants is used to obtain a lower bound on diffusion. Our goal here is to obtain such a bound. In the previous results in QSL we only had one element separated from all others, but what is the relevant space-time distance that bounds correlations when we want to talk about the connected correlation $   c_n\big( A_1(x_1,t_1) , \ldots , A_n(x_n,t_n) \big)$?  We will see that $ c_n\big( A_1(x_1,t_1) , \ldots , A_n(x_n,t_n) \big)$ is controlled by the max-min of the  distances $z_{ij} \coloneqq \dist(A_i(x_i),A_j(x_j))$, whenever $t_i> υ^{-1} \max_i \min_j \{ z_{ij} \}$  with $υ>υ_{LR}$, so that whenever at least one observable is outside the light-cones of all others, the cumulant goes to zero.

For clarity we use the notation $A(x,t) \coloneqq ι_x τ_t A$ for $A\in \mathfrak{U}$. Assuming that $ω \big( A B(x,t) C) - ω(AC) ω(B(x,t))$ is bounded by $1/(\dist(B(x),AC))^p$, we will show that higher order connected correlations (and free cumulants) $ c_n\big( A_1(x_1,t_1)\\ , \ldots , A_n(x_n,t_n) \big)$ are bounded by $1/(max_i \{ min_j \{ z_{ij} \} \})^{(p-rD)}$, where $D$ is the lattice dimension and $r$ the polynomial degree of the dependence of clustering on the support sizes of local observables, see \Cref{defn:sizeable}.  The proof is done in \Cref{section:proof_4}. Note that the assumed three-element cluster property will be true in any state that is clustering in space with rate $1/(1+x)^{p+rD}$, $p>1$, which will be space-like clustering (see \Cref{appendix:space-like}) with rate $1/(1+x)^p$ as a consequence of the Lieb-Robinson bound. This is discussed in \Cref{section:three-element-clustering}.

\begin{theorem} \label{th:n-order-lieb-robinson}
    Consider a short-range quantum lattice $C^*$ dynamical system $(\mathfrak{U},ι,τ)$ and a state  $ω$ for which there exist  $p,r>1$ with $p-rD>1$, such that for every $A,B,C\in \mathfrak{U }_{\rm loc}$,$υ>υ_{LR}$, $x\in \Z^D , t \in -υ^{-1}[ -\dist(B(x),AC), \dist(B(x),AC) ]$:
\begin{equation} \label{eq:three-element-LR-clustering}
   |ω \big( A B(x,t) C) - ω(AC) ω(B(x,t))| \leq  \frac{C_2(B,AC)}{\big(1+ \dist(B(x), AC)\big)^p}   
\end{equation}
where $C_2(A,B)=u\norm{A} \norm{B} P_r( |Λ_A|, |Λ_B|)$ with $u>0$ constant that depends on $υ$, $P_r$ a polynomial of degree $r$ and $|Λ_A |$ the size of the support of $A$.

It follows that for any $n \in \N$ the $n$-th joint cumulant of $A_1,\ldots,A_n \in \mathfrak{U}_{loc}$ satisfies the following clustering property for every $υ > υ_{LR}$:
\begin{equation}  \label{eq:th:n-order-lieb-robinson}
    c_n\big( A_1(x_1,t_1) , \ldots , A_n(x_n,t_n) \big) \leq  \frac{C_n(A_1,\ldots, A_n)}{\big (1+ \max_i \min_j \{\dist(A_i(x_i), A_j(x_j) \} \big)^{p-rD}}
\end{equation}
for $x_1,\ldots,x_n \in \Z^D$, $\displaystyle  t_1,\ldots,t_n \in [-υ^{-1}z+1,υ^{-1}z-1]$, where $ z \coloneqq \\ \max_i \min_j \{\dist(A_i(x_i), A_j(x_j) \}$ and $ C_n(A_1,\ldots, A_n)$ is a function of the $C_2$ and $υ$. This is also true for the free cumulants.
\end{theorem} 

To show \Cref{eq:th:n-order-lieb-robinson} from \Cref{eq:three-element-LR-clustering} we need the following technical lemma. Its proof is discussed in \Cref{appendix:lemma}.
\begin{lem} \label{lemma}
    Consider the assumptions of \Cref{th:n-order-lieb-robinson}. For any $n,m\in \N$ with $m \leq n$, $A_1,\ldots,A_n \in \mathfrak{U}_{\rm loc}$, $x_1,\ldots, x_n \in \Z^D$ and $υ> υ_{LR}$, it follows that 
    \begin{eqnarray}
      \big | ω \big( A_1(x_1,t_1) , \ldots , A_n(x_n,t_n) \big) - ω(A_m(x_m,t_m) ) ω\big( \prod_{j\neq m} A_j(x_j,t_j) \big) \big|\\  
        \leq C^{\prime}_2(A_1,\ldots,A_n) \frac{1}{ (1+ \min_i \{z_{mi} \} )^{p-rD}} 
    \end{eqnarray}
for $\displaystyle  t_1,\ldots,t_n \in υ^{-1}[-\min_{i \neq m} \{ z_{mi} \}, \min_{i \neq m} \{ z_{mi} \}] $, where  $  z_{mi} \coloneqq \dist( A_m(x_m), A_i(x_i) )$.
\end{lem}

If the state $ω$ is exponentially clustering, then it satisfies the Theorem for every $p>1$, thus its higher-order correlations will also be exponentially clustering. The case of long-range interactions will yield a similar result, that will be affected by the degree $a$ of power-law decay.

\subsection{Three element clustering property} \label{section:three-element-clustering}
 
Consider the general set-up introduced in \Cref{section:set-up}. A state $ω$ satisfies the three-element clustering property when  
\begin{equation}
  \lim_i \big(  ω( A α_{g^i}(B) C) - ω(AC) ω(a_{g^i}B \big) =0  ,\, \forall A,B,C \in \mathfrak{U}
\end{equation}
This property is important for proving more general clustering properties of higher order connected correlations, as was the case in \Cref{section:results1}. It is established that in an asymptotically abelian algebra the two-element and three-element clustering properties are equivalent  \cite[Section 4.3.2]{bratteli_operator_1987}. This is shown by considering the limit
\begin{equation}
    \lim_i ω( A [α_{g^i}B,C] ) = 0 
\end{equation}
which is $0$ by asymptotic abelianness. 

Consider now the set-up of QSL with a short-range interaction. We will show that a state satisfying space-like clustering (between two observables)  will satisfy the three-element clustering property with the same rate. Suppose $ω$ is a state that is space-like $q$-clustering for some $q>1$; for $υ>υ_{LR}$, $A,B\in \mathfrak{U}_{\rm loc}$ and any $x \in \Z^D$, $t \in υ^{-1} [-\dist(A(x),B), \dist(A(x),B)]$  we have the bound:
\begin{equation} \label{eq:qsl_spacelike_clustering}
        |ω( A(x,t) B) - ω(A) ω(B) | \leq C_2(A,B) \frac{1}{(1+ \dist(A(x),B))^{q}} 
\end{equation}
See Appendix \ref{appendix:space-like} for more details on how one obtains a space-like clustering bound. By the Lieb-Robinson bound we also get
\begin{equation}
    |ω(A [B(x,t),C)])| \leq \norm{A} \norm{[B(x,t),C)]} \leq L_{B,C} \norm{A}e^{-λ( \dist(B(x),C)- υ_{LR}|t|)} 
\end{equation}
and we also have that
\begin{eqnarray}
    |ω(A [B(x,t),C)])|  &\geq& |ω(A B(x,t) C) - ω(AC)ω(B) |  \\
    &-& |ω(AC B(x,t) ) -ω(AC)ω(B) |
\end{eqnarray}
Combining the above two inequalities:
\begin{eqnarray}
     |ω(A B(x,t) C) - ω(AC)ω(B) | &\leq&  |ω(AC B(x,t) ) -ω(AC)ω(B) | \\
     &+& L_{B,C} \norm{A}e^{-λ( \dist(B(x),C)- υ_{LR}|t|)} 
\end{eqnarray}
and the first term in the right-hand side is bounded by \Cref{eq:qsl_spacelike_clustering}, while the exponential decay is dominated by the polynomial. Therefore we obtain a three-point $q$-clustering property, with the same rate, of the form
\begin{eqnarray}
     |ω(A B(x,t) C) - ω(AC)ω(B) | \leq \frac{ C^{\prime}_2(B,AC) } {(1+\dist(B(x),AC))^{q}}
\end{eqnarray}
for $υ>υ_{LR}$, $A,B,C\in \mathfrak{U}_{\rm loc}$ and any $x \in \Z^D$, $t \in υ^{-1} [-\dist(B(x),AC), \dist(B(x),AC)]$.

\section{Proofs} \label{section:proofs}

\subsection{Proof of  \Cref{th:cumulants_general} and \Cref{th:decay}} \label{section:proof_1}
\begin{proof}
    We prove  \Cref{th:decay}, as  \Cref{th:cumulants_general} follows by choosing $f(g) = 1$.
    We will prove the claim by induction. For $n=2$ the (scaled) second cumulant $f(g^i)κ_2( α_{g^i} A, B) = f(g^i)\,(ω(α_{g^i}( A )B) - ω(α_{g^i}A) ω(B))$ does indeed vanish by the assumption. Suppose \Cref{eq:maindecay} is true for every $m\leq n$, for some $n\in \N$. We want to show that this implies \Cref{eq:maindecay} for $n+1$.
 
    Consider the moments-to-cumulants formula \Cref{eq:moments_to_cumulants}:
    \begin{equation} \label{eq:proof0}
        ω_{n+1}(α_{g^i}A_1,A_2, \cdots, A_{n+1}) = \sum_{π \in NC(n+1)} κ_π (α_{g^i}A_1,\cdots,A_{n+1}).
    \end{equation}
On the r.h.s. the sum $\sum_{π \in NC(n+1)}$ has only one maximal partition of size $1$, the partition $\{1,2,\cdots ,n+1\}$. The maximal partition corresponds to $κ_{n+1}$, which we want to show vanishes. We write
\begin{equation} \label{eq:rhs}
 \begin{array}{*3{>{\displaystyle}lc}p{5cm}}
    r.h.s. &\coloneqq& \sum_{π \in NC(n+1)} κ_π (α_{g^i}A_1,\cdots,A_{n+1}) &=&κ_{n+1} (α_{g^i}A_1,\cdots,A_{n+1})\\ 
    &+& \sum_{π \in NC(n+1), |π|\geq 2} κ_π (α_{g^i}A_1,\cdots,A_{n+1})
    \end{array}
\end{equation}
and then rearrange the terms in \Cref{eq:proof0}:
\begin{equation} \label{eq:proof0.1}
 \begin{array}{*3{>{\displaystyle}lc}p{5cm}}
 && κ_{n+1}(α_{g^i}A_1,\cdots,A_{n+1}) =& ω_{n+1}(α_{g^i}A_1,A_2, \cdots, A_{n+1}) \\
    &-& \sum_{π \in NC(n+1), |π|\geq 2} κ_π (α_{g^i}A_1,\cdots,A_{n+1}).
    \end{array}
\end{equation}          
The set $K=\{π \in NC(n+1), |π|\geq 2\}$ is a union of two disjoint sets: those partitions where the element $1$ is a singleton $K_1=\{π \in NC(n+1), |π|\geq 2 \wedge \{1\}\in π\}$ and those were it is in a block of size at least $2$, $K_2=\{π \in NC(n+1), |π|\geq 2 \wedge (\exists V \in π : 1\in V, |V|\geq 2)\}$.

Hence:
\begin{equation}\label{eq:proof1}
 \begin{array}{*3{>{\displaystyle}lc}p{5cm}}
    \sum_{π \in NC(n+1), |π|\geq 2} κ_π (α_{g^i}A_1,\cdots,A_{n+1}) &=\ \sum_{π\in K_1} κ_π (α_{g^i}A_1,\cdots,A_{n+1}) \\
    &+\ \sum_{π\in K_2} κ_π (α_{g^i}A_1,\cdots,A_{n+1}).
       \end{array} 
\end{equation}
We note that, as a consequence of the general recursion relation \eqref{eq:recursionNC} on non-crossing partitions, we have the recursion relation $K_1= \{ \{ \{1\}\}\cup \pi: \pi\in NC(2,\cdots,n+1)\}$, where $NC(2,\cdots,n+1)$ is the set of non-crossing paritions of $\{2, 3,\cdots, n+1\}$. Hence, as $κ_π$ is completely determined by the $κ_n$, \Cref{eq:multiplicative_free}, we obtain 
\begin{equation} \label{eq:proof2}
 \begin{array}{*3{>{\displaystyle}lc}p{5cm}}
    \sum_{π\in K_1} κ_π (α_{g^i}A_1,\cdots,A_{n+1}) &=& κ_1(α_{g^i}A_1) \sum_{π\in NC(2,\cdots,n+1)}   κ_π (A_2,\cdots,A_{n+1})\\
    &=&ω(α_{g^i}A_1)  \sum_{π\in NC(2,\cdots,n+1)}   κ_π (A_2,\cdots,A_{n+1}).
 \end{array}
\end{equation}
The second term in \Cref{eq:proof1} is:
\begin{equation}
 \begin{array}{*3{>{\displaystyle}lc}p{5cm}} \label{eq:proof1_sum2}
    \sum_{π\in K_2} κ_π (α_{g^i}A_1,\cdots,A_{n+1}) &=& \sum_{π\in K_2} \prod_{V \in π} κ_{|V|} (α_{g^i}A_1,\cdots,A_{n+1}|V). \\
     \end{array}
\end{equation}
Since in every parition  $π \in K_2$ we have $1$ in a block of size at least $2$, every product in the above sum will contain a $V$ with $1\in V$ and $|V|\geq 2$ but $|V|\leq n$, since $|π|\geq 2$. That is, the product will contain a term $κ_m(α_{g^i}A_1, A_{i_{s_1}}, \cdots A_{i_{s_m}})$, $i_{s_1}< \cdots <i_{s_m}$, with $2\leq m+1\leq n$. By the induction hypothesis 
\begin{equation}
    \lim_{i} f(g^i) κ_m(α_{g^i}A_1, A_{i_{s_1}}, \cdots A_{i_{s_m}}) =0 , \forall 2\leq m+1\leq n \text{ and } i_{s_1}< \cdots <i_{s_m}.
\end{equation}
Therefore,
\begin{equation}
    \lim_{i}  f(g^i)\sum_{π\in K_2} κ_π (α_{g^i}A_1,\cdots,A_{n+1}) = 0.
\end{equation}
Thus, taking the limit of \Cref{eq:proof0.1} and using the moments-to-cumulants formula 
\begin{equation} \label{eq:lim_rhs}
 \begin{array}{*3{>{\displaystyle}lc}p{5cm}}
   && \lim_{i}f(g^i) κ_{n+1}(α_{g^i}A_1,\cdots,A_{n+1}) = \lim_i f(g^i)\bigg( ω_{n+1}( α_{g^i} A_1, A_2,\ldots,A_{n+1}) \\
   &-& ω( α_{g^i} A_1)  \sum_{π\in NC(2,\cdots,n+1)}   κ_π (A_2,\cdots,A_{n+1}) \bigg) \\
   &=&  \lim_i f(g^i)\bigg( ω_{n+1}( α_{g^i} A_1, A_2,\ldots,A_{n+1}) - ω( α_{g^i} A_1) ω_{n}(A_2,\ldots,A_{n+1} )\bigg) \\
   &=&\lim_i f(g^i) \bigg( ω\big( α_{g^i} (A_1) \prod_{j=2}^{n+1} A_j\big) - ω( α_{g^i} A_1) ω(\prod_{j=2}^{n+1} A_j)\bigg) \\
   &=& 0
\end{array}
\end{equation}
where the limit is zero by $f(g)$-clustering of the second cumulant between the observables $α_{g^i}A_1$ and $\prod_{i=2}^{n+1} A_i$. This concludes the proof. The proof is identiacal for $ \lim_{i} κ_n( A_1 ,  \ldots, α_{g^i}A_n) = 0 $

If we additionally assume that  $ \lim_{i}  \big(ω(  A \alpha_{g^i}(B) C) -  ω(α_{g^i} B) ω(AC)\big) = 0$, then can show that $ \lim_{i} κ_n( A_1 ,  \ldots, α_{g^i} (A_m), \ldots, A_n) = 0 $, for any $1<m<n$. Follow the same steps as above, the sum over $K_1$ will again cancel out in the limit by the additional clustering property. Each partition in the sum over $K_2$, \Cref{eq:proof1_sum2}, will contain a block $V$ with $m\in V$ and at least one other element
\begin{equation}
        \sum_{π\in K_2} κ_π (A_1 ,  \ldots, \alpha_{g^i} (A_m), \ldots, A_n)= \sum_{π\in K_2} \prod_{V \in π} κ_{|V|} (A_1 ,  \ldots, \alpha_{g^i} (A_m), \ldots, A_n|V).
\end{equation}
And since $m$ can be in any of the first $i\leq m$ positions in $κ_{|V|}$, we have to modify the induction hypothesis so that $ \lim_{i} κ_l( A_1 ,  \ldots, \alpha_{g^i} (A_m), \ldots, A_l) = 0 $ for every $l\leq n$ and all $m<l$, and then show that it's also true for $n+1$. This induction hypothesis implies that the sum over $K_2$ vanishes and concludes the proof. 
\end{proof} 

\subsection{Proof of \Cref{th:clustering_for_groups} } \label{section:proof_3}
The proof is done inductively in $m,n$. We define the proposition $P(m,n)$ as follows, $P(n,m):$
\begin{equation}
    (\forall A_1,  \ldots, A_n \in \mathfrak{U}) [\lim_i κ_n(  α_{g^i} (A_1),  α_{g^i} (A_2),\ldots,  α_{g^i} (A_m), A_{m+1}, \ldots, A_n) =0]
\end{equation}

By \Cref{th:cumulants_general}, $P(1,n)$ is true for every $n\geq 2$. We will prove inductively that $P(n-1,n) \implies P(n,n+1)$, hence $P(n,n+1)$ will be true for all $n$. Afterwards, we will prove that $(\forall m<n)P(m,n) \implies (\forall m<n)P(m,n+1)$ \footnote{This means that if $P(m,n)$ is true for all $m<n$, then $P(m,n+1)$ holds.}.  These two implications show that $P(m,n)$ is true for any $n,m$, with $m<n$, since a base case $P(1,2)$ is true, $P(1,n)$ for all $n$ is true and $P(m,m+1)$  for every $m$. Then all intermediate cases $P(m,n)$ with $1<m <n-1$ follow from $(\forall m<n)P(m,n) \implies (\forall m<n)P(m,n+1)$. The figure bellow illustrates how we navigate the double induction for the proposition $P(m,n)$.
\begin{center}
\renewcommand{\arraystretch}{1.5}
\begin{equation}
\begin{NiceArray}{*{9}{c}}
    (1,2)     &  (1,3)    & (1,4)  & (1,5)  & \ldots   \\
      \cross  & (2,3 )    & (2,4)  & (2,5)   &           \\
      \cross  &   \cross  & (3,4)  & (3,5)   &           \\
      \cross  &   \cross  & \cross &   (4,5)  &      \ldots   \\ 

\CodeAfter
\begin{tikzpicture}
\begin{minipage}{.7\textwidth}
\begin{scope} [->, thick]
\draw (1-1) -- (1-2);
\draw (1-2) -- (1-3);
\draw (1-3) -- (1-4);
{\color{red}\draw (1-1) -- (2-2) ; 
\draw (2-2) -- (3-3) ;
\draw (3-3) -- (4-4) ;
}
{\color{blue}
\draw (2-2.east) to (2-3.west) ;
\draw (1-2.east) to [bend right=45] (2-3.west) ;
\draw (1-1.east) to (2-3.west) ;
}
\end{scope}
\end{minipage}
\end{tikzpicture}
\hspace*{1.2cm}%
\begin{minipage}{.9\textwidth}
$\rightarrow$ \text{by } $P(1,n) \implies P(1,n+1) $  \\
\\
 {\color{blue}$\rightarrow$  \text{by } $(\forall m<n)P(m,n) \implies \\ (\forall m<n)P(m,n+1)$} \\
 \\
{\color{red}$\rightarrow$ \text{by } $P(n-1,n) \implies P(n,n+1) $}
\end{minipage}
\end{NiceArray}
\end{equation} 
\end{center}

In the spirit of the proof of \Cref{th:cumulants_general},  we write
\begin{equation}  \label{eq:proof2_1}
 \begin{array}{*3{>{\displaystyle}lc}p{5cm}}
 && κ_{n}^{(m)} \coloneqq κ_{n}(α_{g^i}A_1, α_{g^i}A_2,\ldots, α_{g^i}A_m, A_{m+1},\ldots,A_{n}) \\ &=& ω_{n}(α_{g^i}A_1, \ldots, A_{n}) 
    - \sum_{π \in NC(n), |π|\geq 2} κ_π (α_{g^i}A_1,\cdots,A_{n})
    \end{array}
\end{equation}      
and we separate $\{π\in NC(n), |π|\geq 2\}$ into two disjoint subsets. The first subset contains all the partitions $π\in NC(n)$ with all $i \in \{1,\ldots,m\}$ belonging to different blocks from every $j\in \{m+1,\ldots,n\}$. The second subset is the complement of the first, where the $i$'s and the $j$'s mix.
\begin{equation}
    K_1 = \{π\in NC(n): |π| \geq 2, \forall V \in π : V \cap M = \emptyset \vee \ V\cap N = \emptyset \}
\end{equation}
where $M= \{1,2,\ldots,m\}$ and $N=\{m+1, \ldots,n\}$, and
\begin{equation}
  K_2= \{π\in NC(n): |π| \geq 2, \exists V \in π : V \cap M \neq \emptyset \wedge \ V\cap N \neq \emptyset \}
\end{equation}
Then \Cref{eq:proof2_1} becomes
\begin{equation}
 \begin{array}{*3{>{\displaystyle}lc}p{5cm}}
    &&κ_{n}^{(m)} =  ω_{n}(α_{g^i}A_1, \ldots, A_{n})   - \sum_{π \in K_1} κ_π (α_{g^i}A_1,\cdots,A_{n}) \\
    &&- \sum_{π \in K_2} κ_π (α_{g^i}A_1,\cdots,A_{n})
    \end{array}
\end{equation}
First, note that $κ_π = \prod_{V \in π} κ_{|V|}$, so the sum over $K_1$ factorises into sums over partitions of $N$ and $M$ (note in particular that because $M$ and $N$ are each composed of contiguous sites, the non-crossing condition indeed factorises):
\begin{equation}
     \sum_{π \in K_1} κ_π = \sum_{π \in NC(1,\ldots,m)} κ_π (α_{g^i}A_1,\ldots ,α_{g^i}A_m) \sum_{π \in NC(m+1,\ldots,n)} κ_π (A_{m+1},\ldots ,A_{n}) 
\end{equation}
and by the moments to cumulants formula this is equal to a product of the respective joint moments:
\begin{equation}
     \sum_{π \in K_1} κ_π = ω(α_{g^i}A_1,\ldots ,α_{g^i}A_m) ω(A_{m+1},\ldots ,A_{n})
\end{equation}
which combined with the term $ ω_{n}(α_{g^i}A_1, \ldots, A_{n})$ will vanish in the limit by two point clustering. 

Using this, let us first prove $P(n,n+1)$,$\forall n \in \N$ by induction; suppose that $P(l-1,l)$ is true for every $l\leq n$. Then, take the limit of $κ^{(n)}_{n+1}$ in \Cref{eq:proof2_1} where the limit of the sum over $K_1$ vanishes, and we are left with $K_2$:
\begin{equation}
    \lim_i κ^{(n)}_{n+1}  =\lim_i\sum_{π \in K_2} \prod_{V \in π} κ_{|V|} (α_{g^i}A_1,\cdots, α_{g^i}A_n, A_{n+1}|V)
\end{equation}
In any partition $π\in K_2$, by definition of $K_2$, the block $V_1$ of $π$ that contains $n+1$ must also contain at least one element of $\{1,\ldots,n\}$, but not all because $π \in K_2$ has at least two blocks. Therefore the product will have a cumulant $κ_l^{(l-1)}$ of some order $l<n+1$ and $l-1$ translated observables. By the induction hypothesis the limit of every term in the sum above must vanish. Therefore $P(n,n+1)$ is true.

Finally, it remains to show $P(m,n) \implies P(m,n+1)$. Consider again by \Cref{eq:proof2_1} the limit of  $κ_{n+1}^{(m)}$ and assume the induction hypothesis that $P(m,n)$ is true for any $l \leq n$ and any $m< l$. The limit for the terms in $K_1$ vanishes, while the limit of the terms in $K_2$ vanishes by the induction hypothesis: the cumulants in the sum are of order less than $n+1$ and at most $m$ translated observables pair with non-translated ones. Therefore $(\forall m <n)P(m,n) \implies P(m,n+1)$, and this concludes the proof.

\subsection{Proof of higher order mean clutering} \label{section:proof_2}

The proof closely follows that of \Cref{th:cumulants_general}. By assumption, for $n=2$ we have:

\begin{equation}
    \lim_i \frac{1}{μ(U_i)} \int_{U_i} κ_2( α_{g}A , B) \,dμ(g) = 0  , \ \forall A,B \in \mathfrak{U}.
\end{equation}
We assume this holds for every $m \leq n$, for some $n >2$, and proceed to prove the theorem by induction. Consider \Cref{eq:proof0.1} and instead of taking its limit we take the limit of the avrage $\lim_i \frac{1}{μ(U_i)} \int_{U_i}$ and follow the same steps as before. We split the sum over $π\in NC(N+1)$, $|π|\geq 2$ into $K_1$ and $K_2$ and use \Cref{eq:proof2}

\begin{equation} 
\renewcommand{\arraystretch}{2.5}
 \begin{array}{*3{>{\displaystyle}l}p{5cm}}
   & \lim_{i}  \frac{1}{μ(U_i)} \int_{U_i} κ_{n+1}(α_{g}A_1,\cdots,A_{n+1}) dμ(g) =\\
    &\lim_i  \frac{1}{μ(U_i)} \int_{U_i} \bigg( ω_{n+1}( α_{g} A_1, A_2,\ldots,A_{n+1})  \\
   &- ω( α_{g} A_1)  \sum_{π\in NC(2,\cdots,n+1)}   κ_π (A_2,\cdots,A_{n+1}) \bigg) \,dμ(g) \\
    &-  \lim_i  \frac{1}{μ(U_i)} \int_{U_i}\sum_{π\in K_2} κ_π (α_{g}A_1,\cdots,A_{n+1}) \,dμ(g) \end{array}
\end{equation}

By the induction hypothesis and the same reasoning as before, the sum over $K_2$ is zero, while the first term is also zero by the induction hypothesis of mean clustering for $n=2$ between $A_1$ and $B= A_2 A_3 \ldots A_{n+1}$.

\subsection{Proof of n-th order Lieb-Robinson bound} \label{section:proof_4}
Consider $c_n(A_1(x_1,t_1),\ldots,A_n(x_n,t_n))$, write $z_{ij}= |x_i-x_j| $ and \\ $z \coloneqq \max_i \min_j \{ z_{ij} \}$. Following the proof of \Cref{th:cumulants_general}, we write the $n-$th cumulant as in \Cref{eq:proof0.1}. We single out the observable that is furthest from the rest. Let $1\leq m,l \leq n$ be such that  the values $i=m,\,j=l$ achieve the max-min of $z_{ij}$ (in particular, $z_{ml} = z$), and suppose $m$ is the one such that $\min_j \{ z_{mj} \} = z$\footnote{This has to be true for at least one of $m$ or $l$, since they achieve the max-min, then for at least one of the two their minimum distance from all others has to be the maximum of all other minimum distances.}. We then proceed to split the partitions with $|π| \geq 2$ into $K_1$ that contains the partitions in which $m$ is a singleton, and $K_2$, as in the original proof:
\begin{equation} \label{eq:proof4_1}
 \begin{array}{*3{>{\displaystyle}lc}p{5cm}}
 && c_{n}(A_1(x_1,t_1),\ldots,A_n(x_n,t_n)) = ω_{n+1}(A_1(x_1,t_1),\ldots,A_n(x_n,t_n)) \\
    &-&  ω(A_m(x_m,t_m))  \sum_{π\in P( \{1,\ldots,n\} \setminus \{m\})}   c_π (A_1(x_1,t_1),\ldots, \hat{A_m}, \ldots,A_n(x_n,t_n)) \\
    &-& \sum_{π\in Κ_2}   c_π (A_1(x_1,t_1),\ldots, A_n(x_n,t_n))
    \end{array}
\end{equation}  
where $K_2=\{π \in P(n), |π|\geq 2 \wedge (\exists V \in π : m\in V, |V|\geq 2)\}$.
The first two terms in the right-hand side will be bounded by the clustering assumption for  $\displaystyle  t_1,\ldots,t_n \in υ^{-1}[-\min_{i \neq m} \{ z_{mi} \}, \min_{i \neq m} \{ z_{mi} \}] $, using \Cref{lemma}, by
\begin{equation}
     \frac{C_2(A_1,\ldots,A_n) }{(1+min_i \{z_{mi}\})^{p-rD}} =\frac{C_2(A_1,\ldots,A_n) }{(1+z)^{p-rD}} 
\end{equation}
and by definition of $m$ we have $min_i \{z_{mi}\} \geq z$.
The terms of sum over $K_2$ are products of cumulants of order at most $n-1$. Consider an arbitrary term $π\in K_2$, with $π = \{ V_1, \ldots , V_{|π|} \}$
\begin{equation} \label{eq:proof4_2}
  \prod_{V_i \in π}  κ_{|V_i|} (A_1(x_1,t_1),\ldots,A_n(x_n,t_n) |V_i ) 
\end{equation}
By definition of $K_2$, for every $π\in K_2$ there is a $V^{\prime}  \in π$ with $|V^{\prime} | \geq 2$ and $m \in V^{\prime} $; $V^{\prime}  = \{i_1 \ldots , i_{|V^{\prime} |} \} \ni m$. Thus, in \Cref{eq:proof4_2} we have a term  
\begin{equation} \label{eq:proof4_3}
   I_{V^{\prime}}\coloneqq κ_{|V^{\prime}|} ( A_{i_1} (x_{i_1},t_{i_1}), \ldots , A_m( x_m,t_m) , \ldots,  A_{i_{|V^{\prime} |}}(x_{i_{|V^{\prime} |}},t_{i_{|V^{\prime} |}}) )
\end{equation}
We assume the theorem is true for $1,2,\ldots, n-1$ and show by induction that it's true for $n$. In \Cref{eq:proof4_3} the cumulant is of order at most $n-1$, hence 
\begin{equation}
\begin{array}{*3{>{\displaystyle}lc}p{5cm}}
     I_{V^{\prime}}&\leq& C_{|V^{\prime}|}(A_{i_1}, \ldots, A_{i{|V^{\prime}|}} )  \frac{1}{ (1+ \max_{i \in V^{\prime}} \min_{j  \in V^{\prime}} \{ z_{ij} \} )^{p-rD}} \\
    &\leq& C_{|V^{\prime}|}(A_{i_1}, \ldots, A_{i{|V^{\prime}|}} )  \frac{1}{ (1+ z)^{p-rD}} .
        \end{array}
\end{equation}
Here we used $\max_{i \in V^{\prime}} \min_{j  \in V^{\prime}} \{ z_{ij} \} \geq \min_{j  \in V^{\prime}} \{ z_{mj} \}\geq \min_{j  \in \{1,\ldots,n\}}\{ z_{mj} \} = z_{ml} = z$.
The rest of the  terms in the product \ref{eq:proof4_2} will give a similar bound:
\begin{equation}
    I_{V_i} \leq C_{|V_i|}(A_j: j \in V_i) \frac{1}{ (1+ \max_{i \in V_i} \min_{j  \in V_i} \{ z_{ij} \} )^{p-rD} } \ \leq C_{|V_i|}(A_i : i \in V_i)  
\end{equation}
where $ C_{|V_i|}(A_j: j \in V_i)= C_{|V_i|}(A_{i_1},\ldots, A_{i_{|V_i|}})$ for the indices in $V_i$. Therefore the term in \Cref{eq:proof4_2} will be bounded by a product of the constants
\begin{equation}
    C_n({A_1,\ldots, A_n} )\coloneqq \prod_{V_i \in π} C_{|V_i|}(A_j : j \in V_i)
\end{equation}
times $\frac{1}{(1+z)^{p-rD}}$, which concludes the proof.

\appendix

\section{Multi-point clustering property} \label{appendix:lemma}
Here we give a proof of \Cref{lemma}, which follows the same ideas as the proofs of space-like clustering from space clustering and the Lieb-Robinson bound in \cite[Theorem 8.5]{doyon_hydrodynamic_2022}, \cite[Appendix C]{ampelogiannis_long-time_2023}. Consider local $A_1,\ldots,A_n \in \mathfrak{U}_{\rm loc}$ and $x_1,\ldots,x_n \in \Z^D$, $t_1,\ldots,t_n \in \R$ and the quantity
\begin{equation} \label{eq:appendixA_S}
     S \coloneqq ω \big( A_1(x_1,t_1)  \ldots  A_n(x_n,t_n) \big) - ω(A_m(x_m,t_m) ) ω\big( \prod_{j\neq m} A_j(x_j,t_j) \big)
\end{equation}
where $j\neq m$ indicates $j=1,\ldots,n$ except $m$. The proof relies on approximating the time evolved operators by local ones, by using the Lieb-Robinson bound, and then applying the assumption \Cref{eq:three-element-LR-clustering}. We use the expression \cite[Corollary 3.1]{sims_lieb-robinson_2011}   for the Lieb-Robnson bound between local $A,B$, there exist $L,υ_{LR}>0$\footnote{L depends on the lattice structure and $υ_{LR}$ on the interaction} independent of $A,B$:
\begin{equation}\label{eq:liebrobinsonbound_detailed}
    \norm{[ τ_t (A) , B]} \leq L \norm{A}\norm{B} \min \{|Λ_A|,|Λ_B|\} \exp{-λ( \dist ( A,B) - υ_{LR}|t|)} 
\end{equation}
We then use \cite[Corollary 4.4]{nachtergaele_quasi-locality_2019} and the Lieb-Robinson bound to obtain  for any local $A\in \mathfrak{U}_{Λ_A}$, supported on $Λ_A \subset Z^D$, an approximation of its time evolution $A(t)$ by a  sequence of local observables $A^\nu(t) \in \mathfrak{U}_{Λ_\nu}$ supported on:
\begin{equation} \label{eq:approximation_support}
    Λ_\nu \coloneqq \cup_{x \in Λ_A}B_{x}(\nu), \, \nu=1,2,3, \dots
\end{equation}
where $B_x(\nu)$ is the $D-$ball of radius $\nu$ around $x$:
\begin{equation} \label{eq:localapproximation}
    \norm{A^\nu(t) - A(t)} \leq  L \norm{A}  |Λ_A|  \exp\{-λ(\nu- υ_{LR}|t|)\} 
\end{equation}
with $\norm{A^{\nu}(t)}=\norm{A(t)}=\norm{A}$.

We approximate all observables $A_i(x_i,t_i)$, except $A_m$, by their local approximations, denoted by $A_{i}^{\nu_i}$ for simplicity:
\begin{equation} 
    \norm{A_{i}^{\nu_i} - A_i(x_i,t_i)} \leq  L \norm{A}  |Λ_A|  \exp\{-λ(\nu- υ_{LR}|t_i|)\} ,\, \nu_i=1,2,\ldots
\end{equation}
for $i \in \{1,2,\ldots, n \} \setminus\{m\}$.

Consider now $μ\in \N$ and the quantity of interest approximated by local observables:
\begin{equation}
  S_μ \coloneqq  ω( \prod_{i=1}^{m-1} A_{i}^{μ} \, A_m(x_m,t_m) \, \prod_{i=m+1}^{n} A_{i}^{μ}) - ω(A_m(x_m,t_m)) ω(\prod_{i\neq m} A_{i}^{μ})
\end{equation}
where $\lim_{μ \to \infty} S_μ=S$ is what we have to bound, in order to prove the Lemma. Now let $\nu\in \N$, that we will specify later, with $\nu < μ$ and write:
\begin{equation} \label{eq:appendixA_1}
\begin{array}{*3{>{\displaystyle}lc}p{5cm}}
  S_μ &=& ω\bigg( \prod_{i=1}^{m-1} ( A_{i}^{μ} + A_{i}^{\nu}- A_{i}^{\nu})   \, A_m(x_m,t_m) \, \prod_{i=m+1}^{n} (A_{i}^{μ}+A_{i}^{\nu}-A_{i}^{\nu} ) \bigg) \\  
  &-& ω(A_m(x_m,t_m)) ω\big(\prod_{i\neq m} (A_{i}^{μ}+A_{i}^{\nu} - A_{i}^{\nu}) \big) \\
  &=& ω( \prod_{i=1}^{m-1} A_{i}^{\nu} \, A_m(x_m,t_m) \, \prod_{i=m+1}^{n} A_{i}^{\nu}) - ω(A_m(x_m,t_m)) ω(\prod_{i\neq m} A_{i}^{\nu}) \\
  &+& ω\big(\sum_{\substack{a_1,\ldots,a_n\in \{0,1\}^n \\ \neq (1,\ldots,1)}}\prod_{i=1}^n ζ(i,a_i)\big)  -  ω(A_m(x_m,t_m)) ω\big(\prod_{i\neq m} (A_{i}^{μ} - A_{i}^{\nu}) \big)  
\end{array}
\end{equation}
where $ζ(m,0)= ζ(m,1)= A_m(x_m,t_m)$, while $ζ(i,0)= A_i^{μ}- A_{i}^{\nu}$ and $ζ(i,1)=A_{i}^{\nu}$ for all $i \neq m$. The sum is over all $n-$tuples $a_1,\ldots,a_n$ with $a_i\in \{0,1\}$, except $a_i=1$ for all $i$, indicating whether the distributive property of the product picks $A_i(x_i,t_i) - A_{i}^{\nu}$ or  $A_i^{\nu}$.  The last two terms in \Cref{eq:appendixA_1} are bounded as:
\begin{equation}
  |  ω(A_m(x_m,t_m)) ω\big(\prod_{i\neq m} (A_{i}^{μ} - A_{i}^{\nu}) \big)  | \leq \norm{A_m} \norm{\prod_{i\neq m} (A_{i}^{μ} - A_{i}^{\nu})}
\end{equation}
and
\begin{equation} \label{eq:appendixA_2}
\begin{array}{*3{>{\displaystyle}lc}p{5cm}}
   \big| \sum_{\substack{a_1,\ldots,a_n\in \{0,1\}^n \\ \neq (1,\ldots,1)}} ω\big(\prod_{i=1}^n ζ(i,c_i)\big) \big| \leq \sum_{\substack{a_1,\ldots,a_n\in \{0,1\}^n \\ \neq (1,\ldots,1)}} \norm{\prod_{i=1}^n ζ(i,c_i)}
\end{array}
\end{equation}
for all $ν,μ$.
These are both  exponentially small by \Cref{eq:localapproximation} whenever $ν> υ_{LR}|t_i|$. In particular \Cref{eq:appendixA_2}, we have  $\norm{ζ(i,1)}=\norm{A_{i}^{\nu}} \leq \norm{A_i}$ and (using $\nu < μ$)
\begin{equation} \label{eq:appendixA_3}
    \norm{ζ(i,0)} = \norm{ A_i^{μ} - A_{i}^{\nu}}   \leq  L \norm{A_i}  |Λ_{A_i}|  \exp\{-λ( \nu - υ_{LR}|t_i|)\} 
\end{equation}
and there is always a term $ζ(i,0)$ in each product since $(a_1,\ldots, a_n) \neq (1,\ldots,1)$. 

The two terms in the third line of \Cref{eq:appendixA_1} are of the form $ω(A B(x,t) C)- ω(B(x,t))ω(C)$, with $A,B,C$ local, hence we can use our main clustering assumption:

\begin{equation}\label{eq:appendixA_T1}
\begin{array}{*3{>{\displaystyle}lc}p{5cm}}
    S_{\nu} \coloneqq |ω\big( \prod_{i=1}^{m-1} A_{i}^{\nu} \,  A_m(x_m,t_m) \prod_{j=m+1}^n  A_{j}^{\nu} \big)- ω(A_m(x_m,t_m) ) ω\big( \prod_{j\neq m} A_{j}^{\nu}\big) | \\ \leq C_2(A_m, \prod_{j\neq m} A_j^{\nu})\frac{1}{\big(1+ \dist(A_m(x_m), \prod_{j\neq m} A_{j}^{\nu})\big)^p} 
     \end{array}
\end{equation}
The constant $C_2(A,B)$ is a polynomial, of degree $r$, of the sizes of the supports of $A,B$. For simplicity we will assume that $C_2(A,B)= u \norm{A} \norm{B} |Λ_A|^r |Λ_B|^r$. The support of $\prod_{j\neq m} A_j^{\nu}$ is the union of the supports of each observable in the product, hence:
\begin{equation} \label{eq:appendixA_estimate1}
     C_2(A_m, \prod_{j\neq m} A_j^{\nu}) \leq u \prod_{j=1}^n \norm{A_j} |Λ_{A_m}|^r \sum_{i\neq m} | Λ_{A_i^{\nu}}|^r
\end{equation}
And by definition of $A_i^{\nu}$, its support given by \Cref{eq:approximation_support}  is
\begin{equation} \label{eq:appendixA_estimate2}
    |Λ_{A_i^{\nu}}| \leq |Λ_{A_i} | \, |B_0(\nu)| \leq B_D |Λ_{A_i}| \nu^D
\end{equation}
Where $|B_0(\nu)|$ the size of the $D-$ball of radius $\nu$, which is a polynomial of degree $D$, hence bounded by $B_D \nu^D$, with $B_D>0$ a constant that depends on the lattice dimension.

The last thing to estimate is $\dist(A_m(x_m), \prod_{j\neq m} A_{j}^{\nu})$. The support of $A_{j}^{\nu}$ is the set that contains all points of the support of $A_j(x_j)$ and $D-$balls of radius $\nu$ around all those points.  Since the support of the product of observables is the union of their supports,  the support of $\prod_{j\neq m} A_{j}^{\nu}$ will be equal to the support of $(\prod_{j\neq m} A_{j}(x_j))^{\nu}$:
\begin{equation}
    \rm{supp}(\prod_{j\neq m} A_{j}^{\nu}) = \bigcup_{j\neq m}  \bigcup_{x \in \supp(A_j(x_j))}B_{x}(\nu) = \bigcup_{x \in \supp(\prod_{j\neq m}A_j(x_j))}B_{x}(\nu) \coloneqq A^{\nu}
\end{equation}
Therefore, as $A^{\nu}$ extends a radius $\nu$ around $\prod_{j\neq m}A_j(x_j)$, a simple geometric argument gives
\begin{equation}
    \dist(A_m(x_m),  \prod_{j \neq m} A_j^{\nu}) \geq \dist(A_m(x_m), \prod_{j \neq m} A_j(x_j)) - \nu 
\end{equation}
Again, since the support of the product is the union of the supports, we get
\begin{equation}\label{eq:appendixA_dist}
    \dist(A_m(x_m),  \prod_{j\neq m} A_j^{\nu}) \geq \min_{j\neq m}\{\dist(A_m(x_m),  A_j(x_j)) \} - \nu 
\end{equation}
With this in mind, we now specify  $\nu$ to be:
\begin{equation}
    \nu = \lfloor ε \min_{i\neq m} \{ \dist (A_m(x_m), A_i(x_i) )\} \rfloor
\end{equation}
for some $0<ε<1$ constant. Note that the estimate \Cref{eq:appendixA_3} now requires
\begin{equation}
    \lfloor ε \min_{i\neq m} \{ \dist (A_m(x_m), A_i(x_i) )\} \rfloor > υ_{LR} |t_j| , \, j=1,\ldots,n
\end{equation}
hence our final bound will be valid for times in the compact set:
\begin{equation}
    t_j \in  [ -  υ^{-1}\min_{i\neq m} \{ \dist (A_m(x_m), A_i(x_i) ) \}-1,  υ^{-1} \min_{i\neq m} \{ \dist (A_m(x_m), A_i (x_i) )\}-1] 
\end{equation}
where $υ= υ_{LR}/ε > υ_{LR}$. 
\begin{remark}
    Note that we have the freedom to choose any $0<ε<1$, and this affects two things. If we choose $ε$ near $1$ we will have a loose clustering bound, but for a large compact set of times, while $ε$ near $0$ yields a tighter bound, but only for short times.
\end{remark}

With this choice of $\nu$, and  using $\lfloor x \rfloor \leq x$,  \Cref{eq:appendixA_dist} gives
\begin{equation} \label{eq:appendixA_estimate3}
\begin{array}{*3{>{\displaystyle}lc}p{5cm}}
   \dist(A_m(x_m),  \prod_{j\neq m} A_j^{\nu}) \geq  (1-ε) \min_{j\neq m} \{ \dist (A_m(x_m), A_j(x_j) )\} 
\end{array}
\end{equation}
We denote $z \coloneqq \min_{j \neq m}\{ \dist(A_m(x_m), A_j(x_j) ) \}$. With this, we now return to \Cref{eq:appendixA_T1} and use \Cref{eq:appendixA_estimate1}, \Cref{eq:appendixA_estimate2} and  \Cref{eq:appendixA_estimate3}:
\begin{equation} \label{eq:appendixA_6}
    S_{\nu} \leq  u \prod_{j=1}^n \norm{A_j} |Λ_{A_m}|^r \sum_{i\neq m} |B_D Λ_{A_i}|^r \frac{ (εz)^{rD} }{(1+(1-ε)z)^p}
\end{equation}
for all $μ \in \N$ with $μ$ large enough.  
The rest of the proof is trivial. To summarize, in \Cref{eq:appendixA_1} the last two terms are bounded exponentially for times $t_i$ in the compact intervals specified above; the exponential bound is dominated by the power-law decay $\frac{1}{(1+ z)^q}$ for any $q>1$. The first two terms in \Cref{eq:appendixA_1} are bounded by $\frac{1}{(1+z)^{p-rD}}$, by \Cref{eq:appendixA_6}. These bounds are uniform in $μ$, hence we can take $μ \to \infty$ in  \Cref{eq:appendixA_1} and conclude the proof.

\section{Space-like clustering bound} \label{appendix:space-like}
Consider  the set-up of QSL with a short-range interaction and a state that is $p$-clustering in space
\begin{equation} \label{eq:qsl_p-clustering}
    |ω( A(x) B) - ω(A) ω(B) | \leq k_2(A,B) \frac{1}{\big(1+\dist(A(x),B)) \big)^p} 
\end{equation}
To obtain clustering for space-time translations, from clustering in space, we need to control the growth of $k_2(A,B)$ with respect to the support sizes of the observables. We call such states sizably clustering:

\begin{defn}[Sizable clustering] \label{defn:sizeable}
    A state $ω$ of a QSL dynamical system $(\mathfrak{U},ι,τ)$ is called $r$-sizably $p$-clustering if it satisfies \Cref{eq:qsl_p-clustering} with
    \begin{equation}
    k_2(A,B) = k \norm{A} \norm{B} |Λ_A|^r |Λ_B|^r, \, k>0, \, r \geq 1
\end{equation}
where $Λ_A$ denotes the support of $A\in \mathfrak{U}_{\rm loc}$, i.e.\ the smallest subset of $\Z^D$ such that $A\in \mathfrak{U}_{Λ_A}$.  
\end{defn}
Using the Lieb-Robinson bound, we can approximate time-evolved observables by local ones \cite{nachtergaele_quasi-locality_2019} and  obtain space-like $q$ clustering for $q=p-rD$. See for example \cite[Theorem 8.5]{doyon_hydrodynamic_2022}. We can obtain for $υ>υ_{LR}$, $A,B\in \mathfrak{U}_{\rm loc}$ and any $x \in \Z^D$, $t \in υ^{-1} [-\dist(A(x),B)+1, \dist(A(x),B)+1]$ we have the bound:
\begin{equation} 
        |ω( A(x,t) B) - ω(A) ω(B) | \leq C_2(A,B) \frac{1}{(1+ \dist(A(x),B))^{p-rD}} 
\end{equation}
Similar proofs are done in \cite[Appendix C]{ampelogiannis_2023_almost}, \cite[Appendix C]{ampelogiannis_long-time_2023} for exponential clustering.

 \section{Defining cumulants by Mobius functions} \label{appendix:mobius}
Consider \Cref{defn:classical_cumulants} for the classical cumulants. The set of partitions $P=\cup P(n)$ is partially ordered by inclusion and one can define a Mobius function over any  locally finite partially ordered set \cite{Speed_1983_Cumulants1}.  It is then noted that the coefficients in \Cref{defn:classical_cumulants} correspond to the value of Mobius function of $P$: 
\begin{equation}
    μ_P(π, \mathds{1}_n) = (-1)^{|π|-1}( |π| - 1)!
\end{equation}
where  $\mathds{1}_n$ is the maximal partition of $\{ 1 ,2 , \ldots, n \}$ \cite{Speed_1983_Cumulants1}. The cumulants of \Cref{defn:classical_cumulants} can then be equivalently defined by the Mobius function of the partition lattice:
\begin{equation}
    c_π (A_1, A_2, \ldots, A_n) \coloneqq \sum_{σ \in P(n), σ \leq π} μ_P(σ,π) ω_σ (A_1, A_2, \ldots, A_n) 
\end{equation}

The free cumulants are then defined similarly, by the Mobius function of the partially ordered set of non-crossing partitions $NC$ (ordered by inclusion):
\begin{equation} \label{eq:classical_cumulants_mobius}
    κ_π (A_1, A_2, \ldots, A_n) \coloneqq \sum_{σ \in NC(n), σ \leq π} μ_{NC}(σ,π) ω_σ (A_1, A_2, \ldots, A_n) 
\end{equation}
It follows that $κ_π$, $π\in NC$, and $c_π$, $π\in P$, form  multiplicative families, determined by the n-th cumulants, which are the ones that correspond to the maximal partition $\mathds{1}_n = \{ \{1,2,\ldots,n\} \}$ of $\{ 1, 2, \ldots,n\}$:
\begin{equation} \label{eq:free_cumulants_mobius}
    κ_n \coloneqq κ_{\mathds{1}_n} , \  c_n \coloneqq c_{\mathds{1}_n}
\end{equation}

In both \Cref{eq:classical_cumulants_mobius} and \Cref{eq:free_cumulants_mobius}, we can apply the respective Mobius inversion \cite{rota_1964_mobius} to obtain the moments-to-cumulants formulae.

\section*{Acknowledgments}
This work has benefited from discussions with Theodoros Tsironis. BD is supported by EPSRC under the grant ``Emergence of hydrodynamics in many-body systems: new rigorous avenues from functional analysis", ref.~EP/W000458/1.  DA is supported by a studentship from EPSRC. 

\bibliographystyle{ieeetr}
\bibliography{references}

\begin{thebibliography}{10}

\bibitem{speicher_2019_notes}
R.~Speicher, ``Lecture notes on "free probability theory",'' 2019.
\newblock arXiv: 1908.08125 [math.OA].

\bibitem{myers_2020_fluctuations}
J.~Myers, M.~J. Bhaseen, R.~J. Harris, and B.~Doyon, ``Transport fluctuations in integrable models out of equilibrium,'' {\em SciPost Physics}, vol.~8, p.~007, 2020.
\newblock Publisher: SciPost.

\bibitem{doyon_2020_fluctuations}
B.~Doyon and J.~Myers, ``Fluctuations in {Ballistic} {Transport} from {Euler} {Hydrodynamics},'' {\em Annales Henri Poincaré}, vol.~21, pp.~255--302, Jan. 2020.

\bibitem{doyon_2019_diffusion}
B.~Doyon, ``Diffusion and {Superdiffusion} from {Hydrodynamic} {Projections},'' {\em Journal of Statistical Physics}, vol.~186, p.~25, Jan. 2022.

\bibitem{garcia_2022_OTOC_chaos_review}
I.~García-Mata, R.~A. Jalabert, and D.~A. Wisniacki, ``Out-of-time-order correlators and quantum chaos,'' {\em arXiv preprint arXiv:2209.07965}, 2022.

\bibitem{bhattacharyya_2022_quantum_chaos}
A.~Bhattacharyya, W.~Chemissany, S.~S. Haque, and B.~Yan, ``Towards the web of quantum chaos diagnostics,'' {\em The European Physical Journal C}, vol.~82, p.~87, Jan. 2022.

\bibitem{zhou_2006_multiparty}
D.~L. Zhou, B.~Zeng, Z.~Xu, and L.~You, ``Multiparty correlation measure based on the cumulant,'' {\em Physical Review A: Atomic, Molecular, and Optical Physics}, vol.~74, p.~052110, Nov. 2006.
\newblock Number of pages: 8 Publisher: American Physical Society.

\bibitem{pappalardi_2017_multi_entanglement}
S.~Pappalardi, A.~Russomanno, A.~Silva, and R.~Fazio, ``Multipartite entanglement after a quantum quench,'' {\em Journal of Statistical Mechanics: Theory and Experiment}, vol.~2017, p.~053104, May 2017.
\newblock Publisher: IOP Publishing and SISSA.

\bibitem{Fowler_2023_cumulant_expansion}
P.~Fowler-Wright, K.~B. Arnardóttir, P.~Kirton, B.~W. Lovett, and J.~Keeling, ``Determining the validity of cumulant expansions for central spin models,'' {\em Physical Review Research}, vol.~5, p.~033148, Sept. 2023.
\newblock Number of pages: 9 Publisher: American Physical Society.

\bibitem{Kramer_2015_generalisedMFT}
S.~Krämer and H.~Ritsch, ``Generalized mean-field approach to simulate the dynamics of large open spin ensembles with long range interactions,'' {\em The European Physical Journal D}, vol.~69, p.~282, Dec. 2015.

\bibitem{Robicheaux_2021_beyond}
F.~Robicheaux and D.~A. Suresh, ``Beyond lowest order mean-field theory for light interacting with atom arrays,'' {\em Physical Review A: Atomic, Molecular, and Optical Physics}, vol.~104, p.~023702, Aug. 2021.
\newblock Number of pages: 12 Publisher: American Physical Society.

\bibitem{kubo_1962_cumulant}
R.~Kubo, ``Generalized {Cumulant} {Expansion} {Method},'' {\em Journal of the Physical Society of Japan}, vol.~17, pp.~1100--1120, July 1962.
\newblock Publisher: The Physical Society of Japan.

\bibitem{fricke_transport_1996}
J.~Fricke, ``Transport {Equations} {Including} {Many}-{Particle} {Correlations} for an {Arbitrary} {Quantum} {System}: {A} {General} {Formalism},'' {\em Annals of Physics}, vol.~252, pp.~479--498, Dec. 1996.

\bibitem{Kira_2008_clusterexp}
M.~Kira and S.~W. Koch, ``Cluster-expansion representation in quantum optics,'' {\em Physical Review A: Atomic, Molecular, and Optical Physics}, vol.~78, p.~022102, Aug. 2008.
\newblock Number of pages: 26 Publisher: American Physical Society.

\bibitem{Sanchez_2020_cumulant}
M.~Sánchez-Barquilla, R.~E.~F. Silva, and J.~Feist, ``Cumulant expansion for the treatment of light–matter interactions in arbitrary material structures,'' {\em The Journal of Chemical Physics}, vol.~152, p.~034108, Jan. 2020.

\bibitem{ampelogiannis_2023_almost}
D.~Ampelogiannis and B.~Doyon, ``Almost {Everywhere} {Ergodicity} in {Quantum} {Lattice} {Models},'' {\em Communications in Mathematical Physics}, vol.~404, pp.~735--768, Dec. 2023.

\bibitem{doyon_hydrodynamic_2022}
B.~Doyon, ``Hydrodynamic {Projections} and the {Emergence} of {Linearised} {Euler} {Equations} in {One}-{Dimensional} {Isolated} {Systems},'' {\em Communications in Mathematical Physics}, vol.~391, pp.~293--356, Apr. 2022.

\bibitem{ampelogiannis_long-time_2023}
D.~Ampelogiannis and B.~Doyon, ``Long-{Time} {Dynamics} in {Quantum} {Spin} {Lattices}: {Ergodicity} and {Hydrodynamic} {Projections} at {All} {Frequencies} and {Wavelengths},'' {\em Annales Henri Poincaré}, May 2023.

\bibitem{doyon_2023_mft}
B.~Doyon, G.~Perfetto, T.~Sasamoto, and T.~Yoshimura, ``Ballistic macroscopic fluctuation theory,'' {\em SciPost Physics}, vol.~15, p.~136, 2023.
\newblock Publisher: SciPost.

\bibitem{voiculescu_1985_Symmetries}
D.~Voiculescu, ``Symmetries of some reduced free product {C}*-algebras,'' in {\em Operator {Algebras} and their {Connections} with {Topology} and {Ergodic} {Theory}} (H.~Araki, C.~C. Moore, S.-V. Stratila, and D.-V. Voiculescu, eds.), (Berlin, Heidelberg), pp.~556--588, Springer Berlin Heidelberg, 1985.

\bibitem{voiculescu_1986_addition}
D.~Voiculescu, ``Addition of certain non-commuting random variables,'' {\em Journal of Functional Analysis}, vol.~66, pp.~323--346, May 1986.

\bibitem{voiculescu_1987_multiplication}
D.~VOICULESCU, ``{MULTIPLICATION} {OF} {CERTAIN} {NON}-{COMMUTING} {RANDOM} {VARIABLES},'' {\em Journal of Operator Theory}, vol.~18, no.~2, pp.~223--235, 1987.
\newblock Publisher: Theta Foundation.

\bibitem{speicher_1994_multiplicative}
R.~Speicher, ``Multiplicative functions on the lattice of non-crossing partitions and free convolution,'' {\em Mathematische Annalen}, vol.~298, pp.~611--628, Jan. 1994.

\bibitem{Pappalardi_2022_ETH_free}
S.~Pappalardi, L.~Foini, and J.~Kurchan, ``Eigenstate thermalization hypothesis and free probability,'' {\em Physical Review Letters}, vol.~129, p.~170603, Oct. 2022.
\newblock Number of pages: 6 Publisher: American Physical Society.

\bibitem{jindal_2024_free}
S.~Jindal and P.~Hosur, ``Generalized free cumulants for quantum chaotic systems,'' 2024.
\newblock arXiv: 2401.13829 [cond-mat.stat-mech].

\bibitem{Ueltschi2003Cluster}
D.~Ueltschi, ``Cluster expansions and correlation functions,'' {\em arXiv: Mathematical Physics}, 2003.

\bibitem{Tran_2017_Lieb_Robinson_npartite}
M.~C. Tran, J.~R. Garrison, Z.-X. Gong, and A.~V. Gorshkov, ``Lieb-{Robinson} bounds on {\textless}span class="nocase"{\textgreater}n{\textless}/span{\textgreater}-partite connected correlation functions,'' {\em Physical Review A: Atomic, Molecular, and Optical Physics}, vol.~96, p.~052334, Nov. 2017.
\newblock Number of pages: 8 Publisher: American Physical Society.

\bibitem{ruelle_statistical_1974}
D.~Ruelle, {\em Statistical mechanics: {Rigorous} results}.
\newblock World Scientific, 1974.

\bibitem{Ruelle_1968_classical_gas}
D.~Ruelle, ``Statistical mechanics of a one-dimensional lattice gas,'' {\em Communications in Mathematical Physics}, vol.~9, pp.~267--278, Dec. 1968.

\bibitem{Araki:1969bj}
H.~Araki, ``Gibbs states of a one dimensional quantum lattice,'' {\em Communications in Mathematical Physics}, vol.~14, pp.~120--157, June 1969.

\bibitem{perez_2023_locality}
D.~Pérez-García and A.~Pérez-Hernández, ``Locality {Estimates} for {Complex} {Time} {Evolution} in {1D},'' {\em Communications in Mathematical Physics}, vol.~399, pp.~929--970, Apr. 2023.

\bibitem{Gibibre_1965_quant_gases}
J.~Ginibre, ``Reduced {Density} {Matrices} of {Quantum} {Gases}. {II}. {Cluster} {Property},'' {\em Journal of Mathematical Physics}, vol.~6, pp.~252--262, Feb. 1965.

\bibitem{park_cluster_1982}
Y.~M. Park, ``The cluster expansion for classical and quantum lattice systems,'' {\em Journal of Statistical Physics}, vol.~27, pp.~553--576, Mar. 1982.

\bibitem{bratteli_operator_1987}
O.~Bratteli and D.~Robinson, {\em Operator {Algebras} and {Quantum} {Statistical} {Mechanics} 1: {C}*- and {W}*-{Algebras}. {Symmetry} {Groups}. {Decomposition} of {States}}.
\newblock Operator {Algebras} and {Quantum} {Statistical} {Mechanics}, Springer Science \& Business Media, 1987.

\bibitem{bratteli_operator_1997}
O.~Bratteli and D.~Robinson, {\em Operator {Algebras} and {Quantum} {Statistical} {Mechanics} 2: {Equilibrium} {States} {Models} in {Quantum} {Statistical} {Mechanics}}.
\newblock Operator {Algebras} and {Quantum} {Statistical} {Mechanics}, Springer Science \& Business Media, 1997.

\bibitem{Araki1975uniqueness}
H.~Araki, ``On uniqueness of {KMS} states of one-dimensional quantum lattice systems,'' {\em Communications in Mathematical Physics}, vol.~44, pp.~1--7, 1975.

\bibitem{Speed_1983_Cumulants1}
T.~P. Speed, ``{CUMULANTS} {AND} {PARTITION} {LATTICES} 1,'' {\em Australian Journal of Statistics}, vol.~25, pp.~378--388, Feb. 1983.
\newblock Publisher: John Wiley \& Sons, Ltd.

\bibitem{greenleaf_1969_invariant}
F.~Greenleaf, ``Invariant means on topological groups and their applications,'' in {\em Van nostrand mathematical studies series, no. 16}, Van Nostrand Reinhold Company, 1969.

\bibitem{nachtergaele_quasi-locality_2019}
B.~Nachtergaele, R.~Sims, and A.~Young, ``Quasi-locality bounds for quantum lattice systems. {I}. {Lieb}-{Robinson} bounds, quasi-local maps, and spectral flow automorphisms,'' {\em Journal of Mathematical Physics}, vol.~60, p.~061101, June 2019.
\newblock Publisher: American Institute of Physics.

\bibitem{Lieb:1972wy}
E.~H. Lieb and D.~W. Robinson, ``The finite group velocity of quantum spin systems,'' {\em Communications in Mathematical Physics}, vol.~28, pp.~251--257, Sept. 1972.

\bibitem{nachtergaele_2006_LR}
B.~Nachtergaele and R.~Sims, ``Lieb-{Robinson} {Bounds} and the {Exponential} {Clustering} {Theorem},'' {\em Communications in Mathematical Physics}, vol.~265, pp.~119--130, July 2006.

\bibitem{Hastings_2006_spectralgap}
M.~B. Hastings and T.~Koma, ``Spectral {Gap} and {Exponential} {Decay} of {Correlations},'' {\em Communications in Mathematical Physics}, vol.~265, pp.~781--804, Aug. 2006.

\bibitem{Chen_2019_finite_scrambling}
C.-F. Chen and A.~Lucas, ``Finite speed of quantum scrambling with long range interactions,'' {\em Physical Review Letters}, vol.~123, p.~250605, Dec. 2019.
\newblock Number of pages: 5 Publisher: American Physical Society.

\bibitem{Kuwahara_2020_linear_LR}
T.~Kuwahara and K.~Saito, ``Strictly linear light cones in long-range interacting systems of arbitrary dimensions,'' {\em Physical Review X}, vol.~10, p.~031010, July 2020.
\newblock Number of pages: 12 Publisher: American Physical Society.

\bibitem{frohlich_properties_2015}
J.~Fröhlich and D.~Ueltschi, ``Some properties of correlations of quantum lattice systems in thermal equilibrium,'' {\em Journal of Mathematical Physics}, vol.~56, p.~053302, May 2015.
\newblock Publisher: American Institute of Physics.

\bibitem{sims_lieb-robinson_2011}
R.~Sims, ``Lieb-{Robinson} {Bounds} and {Quasi}-locality for the {Dynamics} of {Many}-{Body} {Quantum} {Systems},'' in {\em Mathematical {Results} in {Quantum} {Physics}}, pp.~95--106, WORLD SCIENTIFIC, May 2011.

\bibitem{rota_1964_mobius}
G.~C. Rota, ``On the foundations of combinatorial theory {I}. {Theory} of {Möbius} {Functions},'' {\em Zeitschrift für Wahrscheinlichkeitstheorie und Verwandte Gebiete}, vol.~2, pp.~340--368, Jan. 1964.

\end{thebibliography}

\end{document}